\pdfoutput=1
\documentclass[amsmath,amssymb,aps,prb,superscriptaddress,twocolumn,floatfix]{revtex4-2}

\usepackage{graphicx}
\usepackage{wrapfig}
\usepackage{bm}
\usepackage[usenames]{color}
\usepackage[bookmarks=false,linkcolor=NavyBlue,urlcolor=NavyBlue,colorlinks,citecolor=NavyBlue]{hyperref}
\usepackage{subfigure}
\usepackage{multirow}   

\usepackage[dvipsnames]{xcolor}
\definecolor{darkblue}{HTML}{0b3873}

\newcounter{para}
\newcommand{\para}{\par\refstepcounter{para}\textbf{{\color{cyan}[\thepara]}}\space}

\usepackage[normalem]{ulem}

\newcommand{\figref}[2]{\hyperref[#1]{\ref{#1}(#2)}}
\newcommand{\figsref}[2]{\hyperref[#1]{\ref{#1}#2}}

\begin{document}
\let\para\relax
\title{Electron-affinity difference distributions as an organizing principle for superconductivity, enabling the discovery of PtPb$_3$Bi
}

\newcommand{\xCornell}{Department of Physics, Cornell University, Ithaca, NY 14853, USA}
\newcommand{\xCornellCS}{Department of Computer Science, Cornell University, Ithaca, New York 14853, USA}
\newcommand{\xEwha}{Department of Physics, Ewha Womans University, Seoul, South Korea\\\vspace{0.7em}}
\newcommand{\xGoogle}{Google Research, Mountain View, CA, USA}
\newcommand{\xPenn}{Computer and Information Science Department, University of Pennsylvania, Philadelphia, PA 19104, USA}
\newcommand{\xPrinceton}{Department of Chemistry, Princeton University, Princeton, New Jersey 08544, United States}
\newcommand{\xMercerCountyCommunityCollege}{Department of Chemistry, Mercer County Community College, West Windsor, New Jersey 08550, United States}

\newcommand{\equalContribution}{These authors contributed equally to the work.}

\author{Omri~Lesser}
\thanks{\equalContribution}
\affiliation{\xCornell}
\author{Yanjun~Liu} 
\thanks{\equalContribution}
\affiliation{\xCornell}
\author{Natalie~Maus}
\thanks{\equalContribution}
\affiliation{\xPenn}
\author{Aaditya~Panigrahi}
\thanks{\equalContribution}
\affiliation{\xCornell}
\author{Krishnanand~Mallayya}
\affiliation{\xCornell}
\author{Albert~Gong}
\affiliation{\xCornellCS}
\author{Anmol~Kabra}
\affiliation{\xCornellCS}
\author{Scott~B.~Lee}
\affiliation{\xPrinceton}
\author{Sudipta~Chatterjee}
\affiliation{\xPrinceton}
\author{Amira~Merino}
\affiliation{\xPrinceton}
\affiliation{\xMercerCountyCommunityCollege}
\author{Kilian~Q.~Weinberger}
\affiliation{\xCornellCS}
\author{Leslie~M.~Schoop}
\affiliation{\xPrinceton}
\author{Jacob~R.~Gardner}
\affiliation{\xPenn}
\author{Eun-Ah~Kim}
\affiliation{\xCornell}
\affiliation{\xEwha}

\begin{abstract}
Predicting the superconducting transition temperature ($T_c$) from crystal structure and composition remains a central challenge in condensed-matter physics, reflecting the absence of a broadly predictive framework connecting microscopic bonding to macroscopic quantum behavior.
Here, we introduce $\mathcal{GP}$-$T_c$, an interpretable, structure- and chemistry-aware Gaussian process model that enables uncertainty-quantified $T_c$ prediction from experimentally accessible inputs.
By encoding local bonding environments as graphlet histograms, we find that the predictive space collapses to a compact set of descriptors: the distribution of electron-affinity (EA) differences between neighboring atoms, together with interatomic distances and simple elemental features, suffices to predict $T_c$ across disparate superconducting families---identifying an overlooked chemical control parameter that underscores the essential role of local structure beyond composition-only approaches.
Our results demonstrate that the EA differences serves as an accessible window into electronic structure providing a mechanism-agnostic physical basis that captures $T_c$ across conventional and unconventional families, including doped charge transfer insulators.
$\mathcal{GP}$-$T_c$ reproduces the experimentally reported $T_c$ range of the infinite-layer nickelate Nd$_{0.8}$Sr$_{0.2}$NiO$_2$, and we predict and experimentally confirm superconductivity in stoichiometric PtPb$_3$Bi ($T_c \approx 3$~K).
To facilitate broad community use, $\mathcal{GP}$-$T_c$ is made available through a web interface for crystal-structure-based prediction, and the same framework identifies additional high-priority superconducting candidates---including SrNiO$_2$ and K(PRh)$_2$---that provide concrete targets for ongoing and future experimental exploration.
\end{abstract}

\maketitle

\para Superconductors exhibit macroscopic quantum coherence with vanishing electrical resistance, enabling transformative technologies ranging from high-field magnets and power transmission to quantum information platforms. Despite more than a century of research, predicting the superconducting transition temperature $T_c$ from a material’s crystal structure and composition remains a central unsolved problem in condensed-matter physics. Known superconductors span chemically and structurally diverse families with widely varying $T_c$, and no general framework reliably links microscopic bonding and structure to macroscopic superconducting behavior.

\para Microscopic theories such as Bardeen--Cooper--Schrieffer (BCS) theory and its Eliashberg extension~\cite{eliashberg_interactions_1960} provide a quantitative description of phonon-mediated superconductivity when detailed material-specific inputs are available. In practice, however, these inputs are often difficult to compute or measure accurately, and predictive power deteriorates outside narrowly defined regimes. For unconventional superconductors, a broadly predictive microscopic theory remains elusive. As a result, superconductivity discovery has relied heavily on empirical heuristics~\cite{matthias_chapter_1957}, chemical intuition, and serendipity, underscoring the need for complementary approaches that can extract predictive structure--property relationships directly from experimental knowledge.

\para To transcend human-scale heuristics, data-driven approaches have increasingly sought to learn $T_c$ directly from known examples.
However, the first wave of machine-learning (ML) studies was fundamentally limited by the absence of structural information~\cite{SuperCon}.
Relying solely on chemical formulas, deep neural networks and random forests were used to learn
broad statistical trends~\cite{Stanev2018npj,li_critical_2020, Konno2021PRB, Hamidieh2018ComputationalMaterialsScience,Matsumoto2019ApplPhysExpress,Pereti2023npj} but saw limits without structural information.
Although more recent Graph Neural Networks~\cite{choudhary_atomistic_2021, gibson_developing_2025} leverage structural information, their implicit incorporation of this information obscures physical insight. Moreover, their focus on
Eliashberg theory~\cite{choudhary_designing_2022} is only suitable for phononic mechanisms.
The challenge, therefore, is to develop a framework that is both \textit{structure-aware} and \textit{interpretable}---one that can capture the holistic nature of bonding and geometry without being constrained by theoretical priors.

\begin{figure*}[t]
    \centering
    \includegraphics[width=\linewidth]{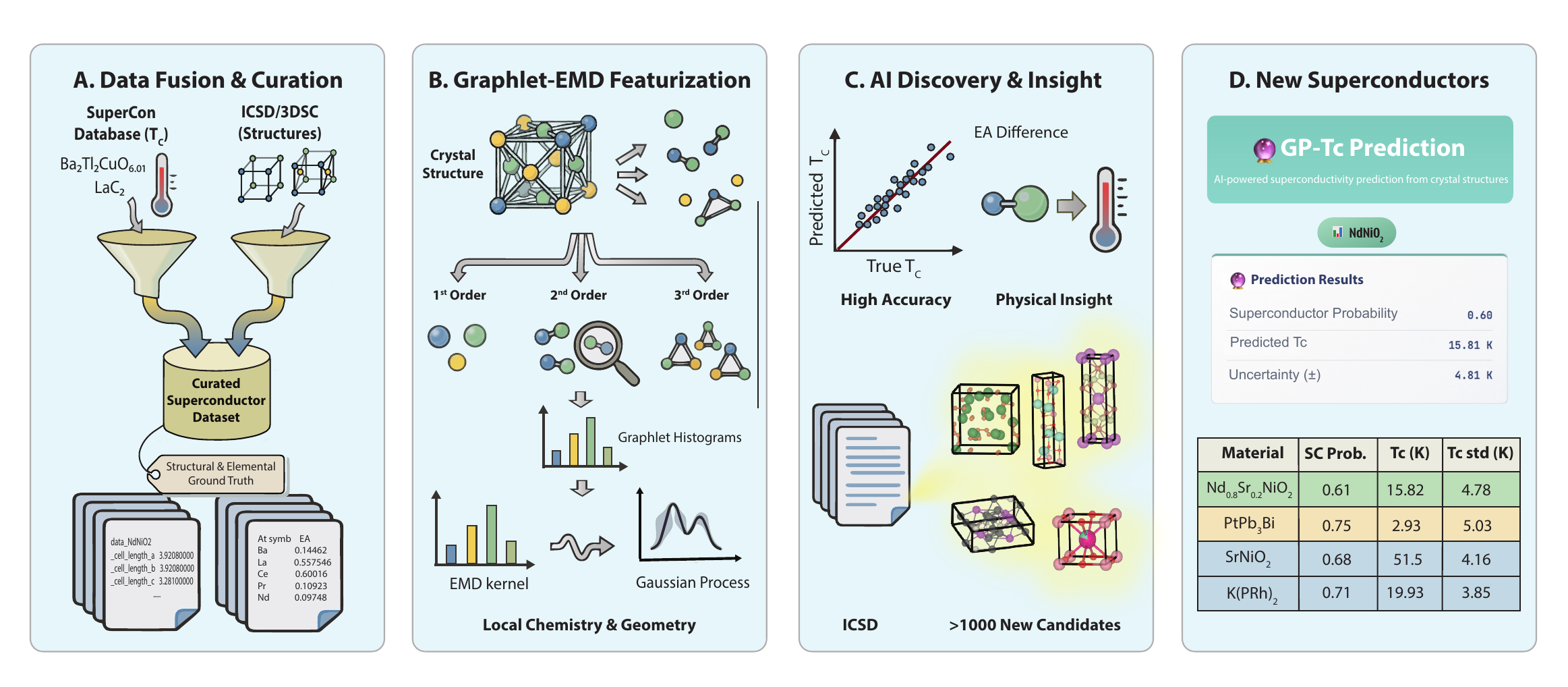}
    \caption{\textbf{Structure-aware machine learning discovers superconductors and reveals organizing principles.}
(a)~Data from known superconductors and tested non-superconductors are curated from the 3DSC database, incorporating crystal structures (from ICSD), measured transition temperatures ($T_c$), and elemental properties.
(b)~Materials are encoded as graphlet histograms capturing local bonding environments at multiple scales (1st order: individual atoms; 2nd order: atom pairs; 3rd order: triplets). Histograms are compared using the Earth Mover's Distance (EMD) to construct a valid probabilistic kernel.
(c)~The Gaussian process model ($\mathcal{GP}$-$T_c$) achieves $R^2=0.93$ while revealing that electron-affinity difference distributions between neighboring atoms dominate prediction. Feature compression from 67 to 4 descriptors exposes compact organizing principles. The framework screens the Inorganic Crystal Structure Database, predicting over 1{,}000 new candidates.
(d)~$\mathcal{GP}$-$T_c$ web interface~\cite{GP_Tc_Webinterface} (top) enables community-wide screening. Experimental validation: Nd$_{0.8}$Sr$_{0.2}$NiO$_2$ $T_c$ reproduced (not in training data); PtPb$_3$Bi discovered ($T_c=2.98$~K vs.\ predicted 2.93~K). High-priority candidates (table) include SrNiO$_2$ ($T_c=51.5$~K).
    }
    \label{fig:intro}
\end{figure*}

\para Here we argue that progress requires combining structure-aware representations with interpretable learning to uncover compact, physically meaningful descriptors of superconductivity.
Rather than pursuing maximal predictive accuracy through increasingly complex models, we seek organizing principles that transparently link local bonding geometry, elemental chemistry, and superconducting behavior in a way that is predictive, insightful, and experimentally actionable.
 To this end, we introduce a structure- and chemistry-aware framework that encodes local bonding environments as graphlet histograms derived from experimentally measured crystal structures, taken from the Inorganic Crystal Structure Database (ICSD), and learns their relationship to superconductivity using an interpretable Gaussian-process model we call $\mathcal{GP}$-$T_c$ (see Fig.~\ref{fig:intro}). This approach enables uncertainty-quantified prediction of $T_c$ while retaining direct access to the features that drive model behavior, allowing systematic identification of the chemical and structural factors most relevant for superconductivity.

\para Importantly, this perspective does not replace microscopic theory but instead provides a complementary route for identifying promising materials and organizing descriptors that can guide both experiment and subsequent first-principles analysis. Applying this framework reveals a striking compression of the predictive feature space: the distribution of electron-affinity differences between neighboring atoms, together with a small number of elemental and structural descriptors, emerges as a central organizing basis for superconductivity across disparate materials families. This insight enables both family-level understanding and experimentally actionable materials discovery, culminating in the prediction and experimental confirmation of a previously unknown stoichiometric superconductor and the identification of multiple high-priority candidates for future exploration [see Fig.~\figref{fig:intro}{d}].


\begin{figure*}[t!]
    \centering
    \includegraphics[width=\linewidth]{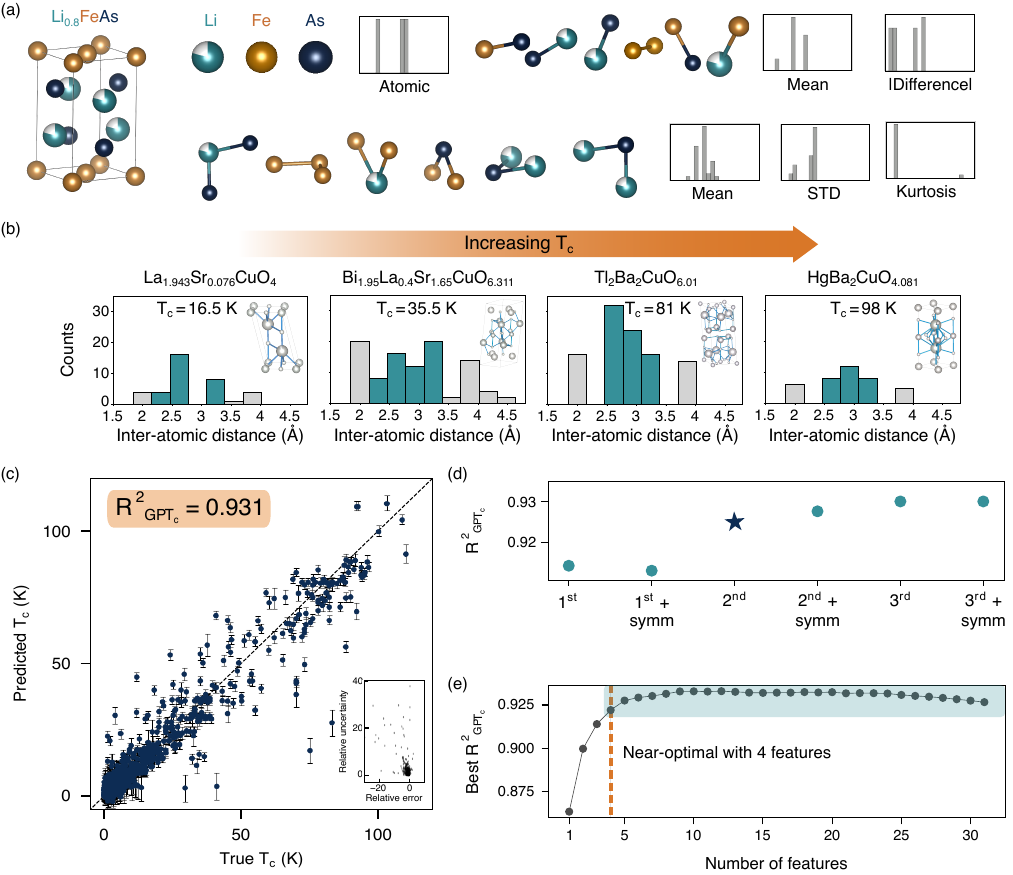}
    \caption{\textbf{Graphlet histograms enable interpretable prediction and reveal dramatic feature compression.}
    (a)~Hierarchical graphlet featurization illustrated with Li$_{0.8}$FeAs: 1st-order graphlets encode individual atoms (elemental properties), 2nd order encode atom pairs (interatomic distances, elemental means/differences), 3rd-order encode triplets (bond angles, higher-order statistics), generating 67 histogram descriptors per material.
    (b)~Interatomic distance histograms for four cuprate superconductors. Blue bars show interlayer distances varying systematically with $T_c$, recovering established correlation between $T_c$ and apical oxygen distances, but within a universally applicable framework that requires no family-specific parameterization.
(c)~Predicted versus experimental $T_c$ for 865 held-out test materials ($R^2=0.931$). Error bars show prediction uncertainties.
The inset shows that the uncertainties generally correlate with prediction errors.
(d)~Feature hierarchy: second-order graphlets significantly improve the performance, while symmetry features provide modest improvement.
(e)~Feature compression: four features achieve near-optimal performance ($R^2=0.922$), revealing compact organizing principles.
   }
    \label{fig:ML}
\end{figure*}

\para A central challenge in predicting superconductivity from crystal structure and composition is constructing a representation that is both physically meaningful and amenable to systematic learning.
We address this through \textit{graphlet histograms}---a systematic featurization that encodes bonding motifs at multiple scales [Fig.~\figref{fig:ML}{a}].
Graphlets describe atomic neighborhoods hierarchically: first-order graphlets characterize individual atoms through 10 elemental properties (Table~S1); second-order graphlets encode neighboring atom pairs through interatomic distances and elemental means and differences (21 features); third-order graphlets extend to atomic triplets, incorporating bond angles (36 features).
In total, 67 histogram features per material encode local chemistry and geometry [Fig.~\figref{fig:intro}{b}] in an architecture-independent manner, enabling interpretability unlike graph neural networks that embed structural information implicitly~\cite{choudhary_atomistic_2021,cerqueira_sampling_2024-1}.

The distributional features are compared using the Earth Mover's Distance (EMD)~\cite{rubner_earth_2000,komiske_metric_2019}, which quantifies the minimal cost to transform one histogram into another and serves as the foundation for constructing a valid Mercer kernel for Gaussian process learning (see Methods).
To capture global symmetry, we construct an 11-dimensional vector by averaging site-specific crystallographic point-group operations over the unit cell (SM Sec.~I).
Combining experimentally measured crystal structures [Fig.~\figref{fig:intro}{a}] with this interpretable probabilistic framework [Fig.~\figref{fig:intro}{c}] enables both accurate prediction and systematic identification of the minimal feature set governing superconductivity.
The resulting model, $\mathcal{GP}$-$T_c$, is made accessible through a web interface [Fig.~\figref{fig:intro}{d}] to enable community-wide materials screening and experimental validation.

\para The predictive utility of graphlet histograms is illustrated by interatomic-distance distributions across cuprate families [Fig.~\figref{fig:ML}{b}]. Decades of cuprate research established that specific structural parameters---such as the apical Cu--O distance---correlate with $T_c$~\cite{rao_relation_1994,pavarini_band-structure_2001,choi_3d_2019}, but such family-specific descriptors lack generalizability. Our histogram features systematically capture these trends without prior knowledge while remaining applicable across all material classes, demonstrating how structure-aware featurization can unify family-specific empirical observations within a broadly transferable framework.

\para We curated a featurized database from 3DSC~\cite{Sommer2023SciData}, which provides experimentally measured crystal structures and $T_c$ values. When multiple structures exist for the same composition, we used the Earth Mover's Distance~\cite{rubner_earth_2000,komiske_metric_2019}---a metric quantifying distributional dissimilarity---to retain only materials with consistent representations (normalized EMD~$<$~0.2), yielding 4,325 materials for training $T_c$ prediction and 5,733 materials for training superconductivity classification (SM Sec.~I.A).

\para To extract interpretable structure--property relationships from graphlet histograms, we employ Gaussian processes (GPs)---a class of probabilistic machine-learning models that provide not only predictions but also quantified uncertainties and direct measures of feature importance~\cite{vivarelli_discovering_1998,rasmussen_gaussian_2006-1,hensman2013gaussian}.
Unlike neural networks that return only point estimates, GPs learn a distribution over functions consistent with the training data, naturally encoding prediction confidence and enabling systematic identification of which features drive model behavior through learned length scales $\ell_n$ (shorter length scales indicate greater predictive power).
This interpretability is essential for our framework: it allows identification of the minimal set of physical descriptors governing superconductivity and quantification of prediction reliability when screening untested materials — capabilities that proved critical for guiding the experimental synthesis pathway for $\mathcal{GP}$-$T_c$ predictions.

\para The central technical challenge in constructing $\mathcal{GP}$-$T_c$ lies in developing a valid GP kernel for histogram features. Standard kernels assume Euclidean feature vectors, but histograms require distance metrics that respect distributional structure. We employ the Earth Mover's Distance (EMD)~\cite{komiske_metric_2019}, which quantifies the minimum cost of transforming one distribution into another. However, EMD does not directly yield a valid Mercer kernel because its Gram matrix is not conditionally negative definite, precluding standard transformations like $\exp(-{\rm EMD}^2)$. We prove (SM Sec.~II.B) that for one-dimensional histograms, the function
\begin{equation}
    K_{\rm EMD}(x_i, x_j) = \sum_{n} w_{n} \mathrm{exp}\left(-\frac{\mathrm{EMD}(h_{i,n}, h_{j,n})}{\ell_{n}}\right),\label{eqn:emd_kernel}
\end{equation}
constitutes a valid kernel, where $x_i = \{h_{i,n}\}$ denotes histogram features for material $i$, and $w_n$, $\ell_n$ are trainable weights and length scales. This result exploits the equivalence between one-dimensional EMD and L1 distance over cumulative distributions. Combining Eq.~\eqref{eqn:emd_kernel} with a radial-basis-function kernel for symmetry features yields the complete $\mathcal{GP}$-$T_c$ model, whose trained length scales $\ell_n$ directly reveal feature importance(see SM Sec.~II for more details).

\begin{figure*}[t!]
    \centering
   \includegraphics[width=\linewidth]{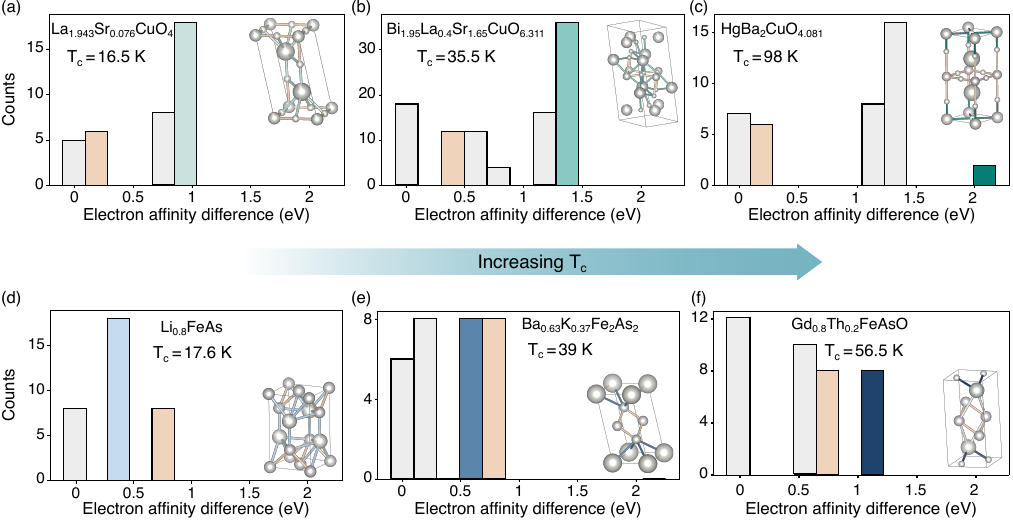}
   \caption{\textbf{Electron affinity difference distributions reveal an organizing principle for superconductivity.}
   (a--c)~Electron affinity (EA) difference histograms for single-layer cuprate superconductors from three families demonstrate systematic trends with $T_c$.
(a)~EA difference histogram for La$_{1.943}$Sr$_{0.076}$CuO$_4$ ($T_c=16.5$~K).
(b)~EA difference histogram for Bi$_{1.95}$La$_{0.4}$Sr$_{1.65}$CuO$_{6.311}$ ($T_c=35.5$~K).
(c)~EA difference histogram for HgBa$_2$CuO$_{4.081}$ ($T_c=98$~K).
(d-f) Iron-based superconductors exhibit the same trend: broader distributions extending to larger EA differences correlate with higher $T_c$.
(d)~EA difference histogram for Li$_{0.8}$FeAs ($T_c=17.6$~K).
(e)~EA difference histogram for Ba$_{0.63}$K$_{0.37}$Fe$_2$As$_2$ ($T_c=39$~K).
(f)~EA difference histogram for Gd$_{0.8}$Th$_{0.2}$FeAsO ($T_c=56.5$~K).
 }
    \label{fig:interpretation}
\end{figure*}

\para $\mathcal{GP}$-$T_c$ achieves state-of-the-art predictive accuracy while offering capabilities unavailable in prior approaches.
Fig.~\figref{fig:ML}{c} shows predicted versus experimental $T_c$ for the held-out test set, yielding $R^2=0.93$---performance comparable to a neural network trained on the same features (SM Sec.~IV) and exceeding prior composition-based models~\cite{Konno2021PRB,Pereti2023npj}.
This robust performance despite using relatively simple machine-learning architectures validates the efficacy of graphlet histogram features: systematic encoding of local bonding environments provides a natural representation that captures the essential chemistry and geometry governing superconductivity across diverse material families.
The test materials span cuprates, iron-based superconductors, heavy fermions, Chevrel phases, and other families, demonstrating broad applicability.
Crucially, a family-resolved linear regression analysis ($T_c^{\text{pred}} = a T_c^{\text{true}} + b$) confirms that the 95\% confidence intervals for the slope $a$ encompass the identity line ($a=1$) across conventional, cuprate, and iron-based families alike (SM Sec.~VII.A). This statistical equivalence demonstrates that the model's predictive power is mechanism-agnostic rather than restricted to conventional phonon-mediated systems.

\para Beyond predictive accuracy, $\mathcal{GP}$-$T_c$ provides two critical advantages for experimental discovery: quantified uncertainty and interpretable feature importance.
As a probabilistic model, $\mathcal{GP}$-$T_c$ inherently estimates prediction confidence --- larger uncertainties generally accompany larger errors [Fig.~\figref{fig:ML}{c}]. Hence, the uncertainties can guide the experimentalists to prioritize synthesis efforts toward high-confidence predictions. Moreover, because the measured $T_c$ is sensitive to sample quality, these uncertainties also guide iterative refinement strategies.
This probabilistic framework inherently quantifies prediction reliability, a critical capability for guiding costly materials discovery efforts.
Feature interpretability emerges through learned length scales: shorter length scales indicate greater predictive influence. Systematic analysis reveals that second-order graphlet features (encoding neighboring atom pairs) provide most of the predictive power [Fig.~\figref{fig:ML}{d}], with first- and third-order features contributing marginally.
This hierarchy highlights the importance of pairwise bonding between neighboring sites, validating our chemistry-informed feature design.

\para The interpretability of $\mathcal{GP}$-$T_c$ enables aggressive feature compression.
Through iterative pruning---sequentially removing the least informative feature and retraining---we identify that just four features achieve near-optimal performance ($R^2=0.922$) [Fig.~\figref{fig:ML}{e}].
To confirm this minimal set, we exhaustively trained $\mathcal{GP}$-$T_c$ on all $\binom{32}{4}=35{,}960$ possible four-feature combinations from the 32 second-order and symmetry descriptors (SM Sec.~V.A).
This dramatic compression from 67 to 4 features suggests that superconductivity, despite its complexity, is governed by surprisingly compact organizing principles — an insight accessible only through the interpretability of the Gaussian process framework, which enables systematic identification of the minimal predictive feature set and opens pathways to physical understanding.

\para The exhaustive search over four-feature combinations reveals a consistent pattern.
Three sets achieve near-optimal performance ($R^2\approx0.92$), all sharing two common descriptors: electron affinity (EA) difference between neighboring atoms and interatomic distance. The third and fourth features vary---periodic table column, valence electron count (total or d-orbital), or atomic weight---but contribute less individually, with notable redundancy among elemental properties (e.g., valence count correlates with periodic table position, as Matthias noted decades ago~\cite{matthias_chapter_1957}).
Critically, all optimal sets employ exclusively second-order graphlet features, confirming that the pairwise bonding character governs $T_c$ prediction.
Symmetry features, though important for classifying whether materials superconduct, prove insignificant for predicting $T_c$ itself (SM Sec.~IV)~\footnote{The symmetry features play a more prominent role in predicting whether a material will superconduct (see SM Sec.~IV).}.
This compressed feature space-where EA difference and interatomic distance provide the core predictive power-enables direct physical interpretation.
As noted in Fig.~\figref{fig:ML}{b}, interatomic distance histograms systematically encode family-specific structural trends (such as apical oxygen distances in cuprates) within a universally applicable framework.
Furthermore, evaluating $\mathcal{GP}$-$T_c$ on the $RE\text{Ba}_2\text{Cu}_3\text{O}_{6+x}$ ($RE$-123) cuprate family demonstrates that graphlet histograms subtly encode off-stoichiometric bonding shifts, reproducing the non-monotonic $T_c$ dome versus oxygen content $x$ without explicit mechanism-specific inputs (SM Sec.~VII.B).

\para The emergence of EA difference as the single most informative feature warrants careful consideration.
Electron affinity quantifies the energy released when a gas-phase atom gains an electron, a directly measurable atomic property.
The EA difference between neighboring sites provides an inexpensive, purely chemical window into local electronic structure without requiring band-structure calculations or extractions. In retrospect, given that the bonding character determines the band structure and degree of correlation, it makes sense that EA difference plays a crucial role in predicting superconducting properties. {For transition-metal oxides}, this connection becomes explicit: the EA difference maps directly onto the charge-transfer energy $\Delta$ in the Zaanen--Sawatzky--Allen framework~\cite{ZaanenSawatzkyAllen1985}. In the cuprates, $\Delta$ directly sets the antiferromagnetic superexchange coupling $J \propto t_{pd}^4 / \Delta^2 (1/U_d + 1/\Delta)$ within the Emery model~\cite{Emery1987,ZhangRice1988}, establishing that EA difference is not distant from spin fluctuations but rather encodes the local bonding character from which magnetic pairing scales emerge (SM Sec.~V.C).
Historically, superconductivity research focused on Pauling electronegativity (EN)---a theoretical, dimensionless scale combining multiple atomic properties~\cite{Luo1987JournalofPhysicsandChemistryofSolids,Jayaprakash1993JournalofPhysicsandChemistryofSolids,rabe_global_1992}---as a proxy for bonding character.
Our analysis reveals that EA difference, despite being overlooked in favor of EN, provides superior predictive power.
This finding is particularly striking because EA is experimentally accessible, whereas EN is derived.

Examining EA difference distributions across material families [Fig.~\figref{fig:interpretation}{a-f}] reveals a clear pattern: within cuprates [panels (a-c)] and iron-based superconductors [panels (d-f)], histogram peaks shift toward larger EA differences as $T_c$ increases across different structural families.
The significance of EA difference distribution suggests that bonding heterogeneity---spatial variations in charge-transfer character between different atom pairs---plays a central role in determining superconducting properties, beyond what average or extremal elemental properties capture.
The distributional nature of this descriptor, which requires histogram representations rather than scalar summaries, underscores why structure-aware featurization is essential for uncovering this organizing principle.

\begin{figure*}[t]
    \centering
    \includegraphics[width=\linewidth]{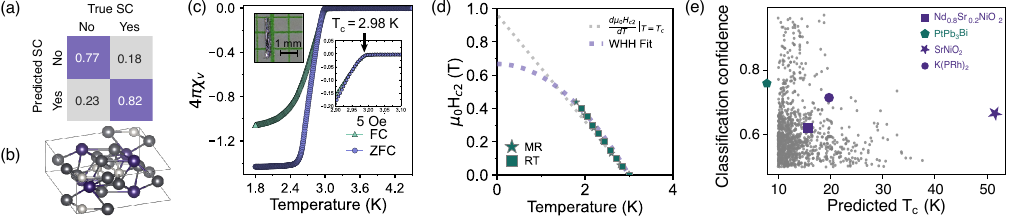}
    \caption{
    \textbf{Experimental validation confirms $\mathcal{GP}$-$T_c$ predictions and identifies high-priority candidates.}
    Classification confidence, combined with the predicted $T_c$, guides experimental discovery.
(a)~Confusion matrix for GP-based superconductivity classifier on held-out test data, achieving 82\% accuracy for superconductors and 77\% for non-superconductors.
(b)~Crystal structure of PtPb$_3$Bi (space group P4$_2$/mnm).
(c)~Magnetic susceptibility versus temperature for PtPb$_3$Bi showing superconducting transition at $T_c=2.98$~K, matching predicted 2.93~K. The field-cooled curve reaches $-1.06$ at 1.8~K, indicating a a complete volume fraction. The inset shows the synthesized sample.
(d)~Electronic transport confirms a bulk superconducting transition. A simplified Werthamer-Helfand-Hohenberg (WHH) fit suggests an upper critical field $\mu_0H_{c2}(0)\approx0.667$~T.
(e)~Predicted new superconductors with $T_c>10$~K after excluding cuprate and iron-based families. Highlighted materials include validated Nd$_{0.8}$Sr$_{0.2}$NiO$_2$ (square, outside training data), high-priority targets SrNiO$_2$ (star, $T_c=51.5$~K), and K(PRh)$_2$ (circle). PtPb$_3$Bi's position (marked in teal on the y-axis) demonstrates that materials with moderate confidence can yield validated discoveries.
    }
    \label{fig:classification}
\end{figure*}

\para We now turn to the classification task of distinguishing superconductors from non-superconductors, which plays a central role in prioritizing experimentally actionable candidates. Rather than treating classification as a standalone benchmark problem, we use it as a reliability filter that complements $\mathcal{GP}$-$T_c$ regression. The classifier assigns each material a superconductivity likelihood based on the same structure-aware representation.
A balanced dataset labeled with ground-truth labels is ideal for training an accurate classifier. However, the labeled data in 3DSC are imbalanced, with a roughly 2.76:1 ratio of superconductors to tested non-superconductors.
Despite this imbalance, predictions correlate strongly with the ground truth [Fig.~\figref{fig:classification}{a}], achieving 82\% accuracy for identifying superconductors and 77\% for non-superconductors, with the asymmetry reflecting the training data composition (SM Sec.~V).
These confidence scores, combined with $\mathcal{GP}$-$T_c$ predictions, enable systematic prioritization of synthesis targets.

\para
We now employ the classifier and $\mathcal{GP}$-$T_c$ on all known inorganic crystals in the ICSD outside of 3DSC using four second-order graphlet features [see Fig.~\figref{fig:intro}{d}] to screen their potential for superconductivity, as most materials have not been tested for superconductivity.
$\mathcal{GP}$-$T_c$ predicts that over 1{,}202 crystalline materials exhibit $T_c>10~{\rm K}$ outside of 3DSC.
Through an exhaustive, AI-based literature search and verification, we confirm that 1{,}032 materials would be new superconductors if experimentally verified and provide what is known about those materials (see SM Sec.~VI and the associated database).
Two experimental verifications of the predictions bolster our confidence: the reproduction of $T_c$ for Nd$_{0.8}$Sr$_{0.2}$NiO$_2$ in the $9$--$15K$ range and our own synthesis and verification of PtPb$_3$Bi.

\para
The SuperCon database largely preceded the 2018 discovery of superconductivity in the infinite-layer nickelates, and consequently, 3DSC contains no nickelate superconductors, including Nd$_{0.8}$Sr$_{0.2}$NiO$_2$~\cite{li_superconductivity_2019-1}.
In this sense, Nd$_{0.8}$Sr$_{0.2}$NiO$_2$ provides a stringent family-level test for our structure-aware framework: $\mathcal{GP}$-$T_c$ predicts $T_c$ in the experimentally reported $9$--$15$~K range, and its uncertainty naturally accommodates sample-to-sample variability. Note the prediction assumes homogeneous Sr doping and uniform, single-phase infinite-layer structure, as Ref.~\cite{li_superconductivity_2019-1} found necessary in the reported thin-film superconductivity.
The classification confidence is 0.606, reflecting the out-of-distribution nature of the material.

\para To demonstrate prospective experimental discovery---rather than reproduction of known results---we selected a candidate that was (i) assigned a high superconductivity classification confidence and (ii) experimentally accessible as a stoichiometric ternary compound with readily available elements (SM Sec.~VI).
PtPb$_3$Bi [see Fig~\figref{fig:classification}{b}] emerged as such a high-confidence target with classification confidence 0.756, with $\mathcal{GP}$-$T_c$ predicting $T_c \approx 2.93$~K.
We synthesized PtPb$_3$Bi in its pure phase (SM Sec.~VI.D) and performed bulk magnetic susceptibility [Fig.~\figref{fig:classification}{c}] and electronic transport [Fig.~\figref{fig:classification}{d}] measurements, which reveal a superconducting transition at $T_c = 2.98$~K---in excellent agreement with the $\mathcal{GP}$-$T_c$ prediction.
The field-cooled susceptibility reaches $-1.06$ at 1.8~K, confirming a complete superconducting volume fraction. A simplified Werthamer-Helfand-Hohenberg (WHH) fit, $\mu_0H_{c2}(0) = -0.693*T_c*\frac{d\mu_0H_{c2}}{dT}|_{_{T = T_c}}$, suggests an upper critical field limit of 0.667~T, far below the Pauli limit, $H_P\approx1.86\times3.0\approx5.6~T$, consistent with an orbital-pair breaking mechanism.
Temperature- and frequency-dependent $I-V$ curves, as well as field-dependent magnetization are all consistent with type-II behavior, characteristic of multi-element intermetallics and indicative of bulk superconductivity (see SM Sec.~VI).
Interestingly, PtPb$_3$Bi forms a unique structure type, hosting face-sharing dimerized [Pt$_2$Pb$_8$Bi$_4$] bicapped trigonal prismatic units, stacked along the c-axis forming a quasi-one-dimensional structure type (see SM Sec.~VI).
To the best of our knowledge, no other variants have been reported among superconductors; this unique structure type would have been overlooked by structure-motif-based or symmetry-constrained search strategies, demonstrating $\mathcal{GP}$-$T_c$'s ability to discover materials beyond theoretical bias.

\para Beyond the experimental validation of PtPb$_3$Bi, our screening of the Inorganic Crystal Structure Database (ICSD) identifies numerous high-priority candidates for experimental investigation.
To facilitate community access, we deployed $\mathcal{GP}$-$T_c$ through an interactive web interface.
Figure~\figref{fig:classification}{e} displays the screening landscape after excluding cuprate and iron-based families, with each point representing a predicted material, positioned by its predicted $T_c$ ($x$-axis) and classification confidence ($y$-axis). It reveals that the high-$T_c$, high-confidence upper-right quadrant contains the most compelling experimental targets.
Among the 1{,}032 predicted new superconductors with $T_c>10$~K (see SM Sec.~VI and associated database), several particularly intriguing examples emerge.
Most notably, SrNiO$_2$ receives a predicted $T_c=51.5$~K (uncertainty $\pm4.16$~K, classification confidence 0.67)---substantially higher than any known nickelate superconductor.
This prediction is especially tantalizing given that Sr doping enables superconductivity in Nd$_{0.8}$Sr$_{0.2}$NiO$_2$ by introducing holes into the insulating parent NdNiO$_2$.
Little is known about SrNiO$_2$, with only one paper reporting the preparation of single crystals of SrNiO$_2$ \cite{Pausch_preparation_1976}; we used the reported tetragonal CmCm structure in our prediction. \footnote{Wang et~al. predicted a Hund's metal at 100~K for the infinite-layer $P4/mmm$ structure. However, the (La,Sr)NiO$_2$ system, which was also predicted to be a Hund's metal, has recently been confirmed to be superconducting at 9~K~\cite{osada_Superconductivity_2021}.}
The high predicted $T_c$ suggests that the stoichiometric compound and its electron-doped variants warrant systematic investigation.
Other high-confidence predictions include K(PRh)$_2$ [highlighted in Fig.~\figref{fig:intro}{d}], providing concrete experimental targets across diverse chemical families.

\para To summarize, we introduced a structure-aware and interpretable framework for superconductivity prediction that explicitly integrates elemental chemistry with local bonding geometry derived from experimentally measured crystal structures. By encoding materials using graphlet histograms and learning within a Gaussian process model ($\mathcal{GP}$-$T_c$), we achieve accurate, uncertainty-quantified predictions of both superconductivity and $T_c$ while retaining direct access to the physical features that govern model behavior. This design choice enables not only predictive performance but also the extraction of compact, chemically meaningful organizing principles.

\para A central outcome of this approach is the dramatic compression of the predictive feature space. We find that the distribution of electron-affinity differences between neighboring atoms, together with a small number of elemental and structural descriptors, suffices to achieve near-optimal performance in $T_c$ prediction. That such a readily available and directly measurable quantity emerges as the most informative feature highlights the central role of local bonding character in superconductivity. This result shifts attention away from opaque high-dimensional descriptors toward chemically interpretable variables that can be directly connected to bonding physics and materials design.

\para Beyond insight, $\mathcal{GP}$-$T_c$ demonstrates reliable experimental actionability. We reproduce the superconducting behavior of the recently established infinite-layer nickelate family and predict and experimentally confirm superconductivity in stoichiometric PtPb$_3$Bi, establishing a new superconductor. At the same time, the framework identifies multiple additional high-priority candidates, including SrNiO$_2$ with a predicted $ T_c$ of 51.5~K, providing concrete targets for experimental exploration. By making $\mathcal{GP}$-$T_c$ accessible through a web interface \cite{GP_Tc_Webinterface} that enables structure-based prediction from standard CIF files, we aim to lower the barrier to community-wide exploration. More broadly, because the featurization relies only on fundamental elemental and structural information, the strategy developed here is readily extensible to other material properties. Being physically interpretable rather than opaque learned embeddings, these histograms could similarly serve as generation targets for crystal-structure models (e.g., DiffCSP, MatterGen, SCIGEN) \cite{jiao_crystal_2023,zeni_mattergen_2025,okabe_structural_2025}, guiding candidate generation toward chemically meaningful bonding environments. Such guided generation would open a path toward interpretable, data-driven discovery across quantum materials.

{\bf Author Contributions:}
E.-A.K. and J.G. planned the machine learning strategy and guided the research activities.
K.M. devised the graphlet expansion and the EMD metric.
N.M. built the EMD kernel and trained the GP models.
O.L. carried out NN-based studies and led the interpretation efforts.
Y.L. designed the symmetry features and analyzed ML results.
Y.L. and A.P. wrote the code for graphlet and symmetry feature generation and curated the database. A.P. implemented the prediction pipeline for the ICSD screening. A.P. and O.L. compiled the predicted superconducting candidates. A.P. implemented the $\mathcal{GP}$-$T_c$ prediction web interface. A.G. and A.K. developed the literature-search tools used to analyze the predicted superconducting materials.
K.W. advised an LLM-based comprehensive literature search to identify precedents for $\mathcal{GP}$-$T_c$ predictions.
A.G. and A.K. designed and carried out the LLM-based comprehensive literature search to identify precedents for $\mathcal{GP}$-$T_c$ predictions.
L.S. advised the choice of atomistic features, the classification experiments, the choice of PtPb$_3$Bi as the first test case material, its synthesis, its characterization, and its measurements.
S.L., S.C., and A.M. synthesized, characterized, and measured PtPb$_3$Bi.

{\bf Acknowledgements:} We are grateful to Andrei Bernevig, Timo Sommer, Pascal Friederich, J\"{o}rg Schmalian, Ichiro Takeuchi, and Steve Kivelson for helpful discussions.
E.-A.K., K.W., and L.S. are supported by the NSF through the AI Research Institutes program Award No.~DMR-2433348 and by the grant OAC-2118310.
Y.L. and E.-A.K. were supported in part by the MURI grant FA9550-21-1-0429. K.M. was supported by Eric and Wendy Schmidt AI in Science Postdoctoral Fellowship: a Schmidt Futures program.
The computation was performed on a high-performance computing cluster supported by the Gordon and Betty Moore Foundation’s EPiQS Initiative, Grant GBMF10436 to E.-A.K., and by the MURI grant FA9550-21-1-0429.
K.M., Y.L., and E.-A.K. are supported by the U.S. Department of Energy, Office of Science, Basic Energy Sciences, Materials Sciences and Engineering Division.
O.L. and E.-A.K. are supported by the U.S. Department of Energy through Award Number DE-SC0023905.
O.L. is also supported by a Bethe-KIC postdoctoral fellowship at Cornell University.
J.G. was supported by NSF grants DBI-2400135 and IIS-2145644. N.M. was supported by an NSF Graduate Research Fellowship.
S.L. and S.C. are supported by the Air Force Office of Scientific Research under award number FA9550-25-1-0177.
A.M. was supported by an NSF CAREER grant (DMR-2144295) to L.S.

{\bf Data and code availability: } Featurized training data supporting the findings of this study, provided as graphlet histograms, together with code implementing the full data pipeline, including crystal structure graphlet featurization, Gaussian process model training, and prediction, are available at upon request from the authors. Raw crystal structure inputs are subject to licensing restrictions from the Inorganic Crystal Structure Database (ICSD) and cannot be redistributed.
\bibliography{Refs}

\begin{thebibliography}{66}
\bibitem [{GP_()}]{GP_Tc_Webinterface}%
  \BibitemOpen
  \href@noop {} {\bibinfo {title} {Web-based interface access is available upon
  request from the authors.}}\BibitemShut {Stop}%
\end{thebibliography}


\clearpage
\onecolumngrid

\begin{center}
\large{\textbf{Supplementary Material for\\*[0.5em]``Electron-affinity difference distributions as an organizing principle for superconductivity, enabling the discovery of PtPb$_3$Bi''}}
\end{center}

\setcounter{equation}{0}
\renewcommand{\theequation}{S\arabic{equation}}
\setcounter{figure}{0}
\renewcommand{\thefigure}{S\arabic{figure}}
\setcounter{table}{0}
\renewcommand{\thetable}{S\arabic{table}}
\setcounter{section}{0}

\tableofcontents

\section{Data and features}
In most ML works focusing on superconductivity, the data came from the SuperCon database, which tabulates the experimental measurements of superconductivity for $>$16000 materials and their references~\cite{Hamidieh2018ComputationalMaterialsScience,Stanev2018npj,Zeng2019npj,Matsumoto2019ApplPhysExpress,Roter2020PhysicaC,Le2020IEEE,Konno2021PRB,Pereti2023npj,SuperCon}. For ML studies, the applicable data in SuperCon are only the chemical formula and critical temperatures of the materials. Furthermore, some materials have multiple entries reported by different references with varying reported $T_c$ values. The lack of structural information restricted previous works to composition-only features, eliminating the possibility for the ML algorithms to uncover structure-related mechanisms of the problem. Here we describe a way to leverage more information about the materials, using an enhanced database that recently became available.

\subsection{3DSC down selection}

In this work, we use the 3DSC database~\cite{Sommer2023SciData}, which pairs experimentally measured critical temperatures of superconducting and non-superconducting materials with crystal structures from the Materials Project (MP)~\cite{MP} or the Inorganic Crystal Structure Database (ICSD)~\cite{ICSDweb,ICSDpapaer}, applying artificial modifications when necessary. 3DSC thus provides two datasets, 3DSC$_{\mathrm{MP}}$ and 3DSC$_{\mathrm{ICSD}}$; we focus on the latter, which relies primarily on experimental structural data rather than on first-principles calculations. 3DSC$_{\mathrm{ICSD}}$ covers approximately 57$\%$ of the SuperCon entries (9{,}150 materials). However, many materials are matched with multiple CIFs whose recorded crystal structures are not necessarily identical. Two main sources of structural variation exist for a single chemical formula. First, the material may exhibit polymorphism. Second, many SuperCon compositions do not have exact matches in the ICSD. For such cases, 3DSC uses ICSD entries whose formulas pass its own similarity threshold as parent structures and generates candidate CIFs by synthetically doping them to match the target SuperCon composition. This procedure changes site occupancies while retaining the parent atomic coordinates. Since our featurization encodes structural information, and 3DSC provides both directly matched and synthetically doped CIFs as candidate structures, we develop a quantitative criterion to filter the materials from 3DSC$_\mathrm{ICSD}$ that we can reliably work with.

Our selection criterion is based on graphlet features and Earth Mover's Distances (EMD), both of which are detailed in the following sections. After graphlet featurization, we obtained valid histogram features for 76{,}037 CIFs spanning 8{,}781 materials, of which 57{,}274 CIFs belong to 6{,}463 superconductors with non-zero $T_{c}$. For regression ($T_{c}$ prediction), we include only superconductors. We begin with the 2{,}033 superconductors represented by a unique CIF. To enlarge the dataset, we also seek to include superconductors that are associated with multiple CIFs. Two primary reasons underlie this multiplicity: (1) independent references may report the same crystal structure, producing multiple but nearly identical CIFs; and (2) the material may exhibit polymorphism, or the CIFs may be doped from different parent structures through the 3DSC algorithm, yielding genuinely distinct CIFs. This second case is especially common among cuprates, where many doped compositions have reported $T_c$ values but no directly matching ICSD entries. Filtering out materials represented by mutually inconsistent CIFs therefore reduces the risk that the model learns artifacts of synthetic structural proxies rather than genuine structure--property relationships.

We assess the mutual similarity of a material's CIFs by computing the EMD between their histogram features. After graphlet featurization, each CIF is represented by 10 first-order and 21 second-order histogram features. We first computed EMDs for every pair among the 2{,}033 unique-CIF superconductors, yielding 2{,}065{,}528 EMD values per feature dimension. For each dimension, we sorted these values and defined the 1st percentile as the similarity threshold. Two CIFs are deemed similar if their EMDs in all 31 feature dimensions fall below the respective thresholds. We test the EMD-based similarity test on the set of 2{,}033 unique-CIF superconductors to ensure that the similarity test indeed distinguishes CIFs associated with distinct materials. For this, we evaluated EMD among all possible pairs of CIFs in this set, $\binom{2,033}{2} = 2,065,528$ pairs in total and confirmed that 99.96$\%$ of distinct-material pairs were correctly classified as dissimilar. This benchmarking gave us confidence on our EMD-based similarity test between a pair of CIFs.

We then applied the same graphlet EMD based similarity test to superconductors with multiple CIFs. If all pairwise comparisons among a material's CIFs satisfied the similarity condition, we retained that material and selected one of its CIFs at random, as they were deemed sufficiently alike. This procedure added 2{,}292 superconductors, bringing the total to 4{,}325. We applied the same procedure to the 2{,}318 non-superconductors with 18{,}763 valid CIFs, retaining 1{,}527 for classification. The final classification dataset combines these non-superconductors with the retained superconductors, after removing a small number of heavily oversampled variants from the superconducting class, yielding 4{,}206 superconductors and 1{,}527 non-superconductors (5{,}733 materials in total).

To summarize the down selection, we classify materials according to the CIFs that 3DSC supplies. An \emph{experimental} CIF comes directly from an ICSD entry whose formula matches the SuperCon composition after normalization; the 3DSC doping algorithm never modifies it. A \emph{synthetically doped} CIF is instead produced by the doping algorithm, which alters the site occupancies of an ICSD parent whose composition is similar but not identical. We label a material ``\emph{Experimental-only}'' when all of its CIFs are experimental, ``\emph{Synthetic-only}'' when all are synthetically doped, and ``\emph{Mixed}'' when both types are present. A mixed material therefore always carries multiple CIFs. We further distinguish single-CIF from multi-CIF materials, and stoichiometric (integer-formula) from doped (non-integer) compositions. This grouping serves only to summarize the down selection; the consistency criterion itself compares all CIFs of a given material jointly and never separates experimental from synthetically doped CIFs.

Figure~\ref{fig:branch} traces how the five CIF-evidence categories split into included and dropped materials, and Table~\ref{tab:downselection_summary_streamlined} gives the counts for each branch. The consistency criterion prunes multi-CIF materials whose CIFs disagree. Mixed materials are pruned most often, because a directly matched experimental structure rarely agrees with structures doped from different parents. Materials with multiple synthetically doped CIFs also have a high drop rate, since each CIF is typically doped from a distinct parent. A small number of stoichiometric materials carry only synthetically doped CIFs, even though an integer-formula compound would normally have an exact ICSD match. This situation arises either because the ICSD records the same compound under a different normalization, formula-unit convention, or partial-occupancy setting---causing the matching step to miss the exact formula and route the entry through the doping algorithm, which returns the original structure nearly unchanged---or because the bulk ICSD download omits the exact entry altogether, leaving only doped approximations from nearby parents.

\begin{center}
\begin{minipage}{0.95\textwidth}
\centering
\refstepcounter{figure}\label{fig:branch}
\includegraphics[width=\textwidth]{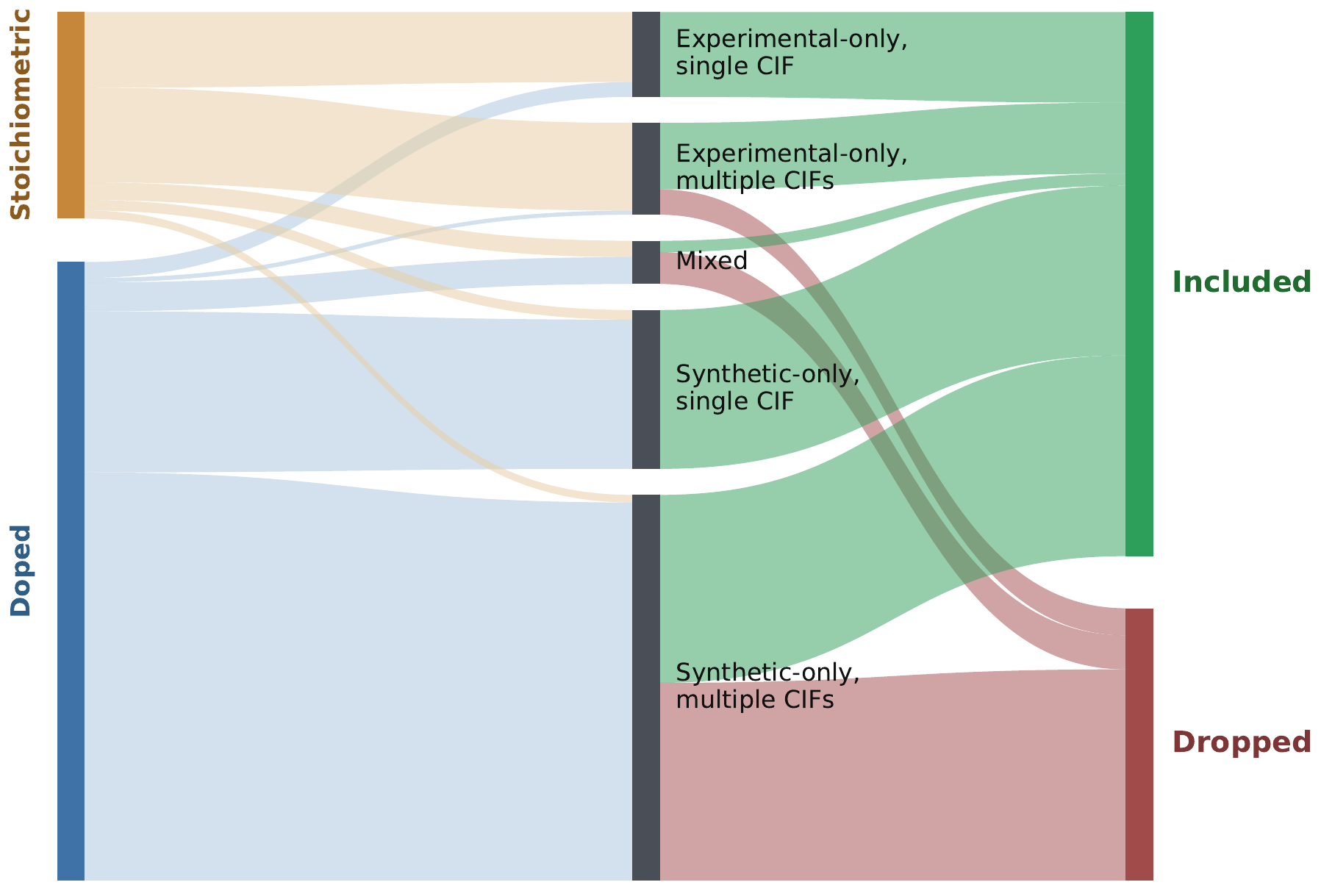}
\parbox{\textwidth}{\textbf{FIG.~\thefigure.} Down-selection outcome for the combined classification set, shown as a three-stage flow. Materials are first divided into stoichiometric and doped compositions, then into the five CIF-evidence categories (experimental-only single, experimental-only multiple, mixed, synthetic-only single, synthetic-only multiple), and finally into included and dropped. Band widths are proportional to the number of materials. All single-CIF materials are retained, whereas multi-CIF materials---especially mixed and synthetic-only multi-CIF materials---lose a large share.}
\end{minipage}
\end{center}
\vspace{0.5em}

\begin{center}
\begin{minipage}{0.95\textwidth}
\centering
\refstepcounter{table}\label{tab:downselection_summary_streamlined}
\parbox{\textwidth}{\textbf{TABLE~\thetable.} Down-selection outcome for the superconducting (SC) candidates used for $T_c$ regression and the non-superconducting (non-SC) candidates with valid graphlet features, by stoichiometry, CIF evidence, and CIF multiplicity. Experimental CIFs are original ICSD entries that match the normalized SuperCon composition, while synthetically doped CIFs are produced by the 3DSC doping algorithm from an ICSD parent of similar but non-matching composition. A mixed material has both experimental and synthetically doped CIFs and therefore always appears in the multiple-CIF column.}
\vspace{-0.6em}
\begin{tabular}{lll@{\hspace{0.8em}}r@{\hspace{0.8em}}r@{\hspace{0.8em}}r@{\hspace{0.8em}}r}
\hline\hline
 & & & \multicolumn{2}{c}{SC} & \multicolumn{2}{c}{non-SC} \\
\cline{4-7}
Composition & CIF evidence & CIF count & kept & dropped & kept & dropped \\
\hline
\multirow{5}{*}{Stoichiometric}
 & Experimental only & single   & $423$ & $0$   & $381$ & $0$ \\
 & Experimental only & multiple & $337$ & $152$ & $391$ & $128$ \\
 & Mixed             & multiple & $8$   & $91$  & $2$   & $87$ \\
 & Synthetic only    & single   & $66$  & $0$   & $45$  & $0$ \\
 & Synthetic only    & multiple & $29$  & $17$  & $22$  & $19$ \\
\cline{2-7}
 & \multicolumn{2}{l}{Subtotal} & $863$ & $260$ & $841$ & $234$ \\
\hline
\multirow{5}{*}{Doped}
 & Experimental only & single   & $136$     & $0$     & $34$  & $0$ \\
 & Experimental only & multiple & $28$      & $8$     & $8$   & $3$ \\
 & Mixed             & multiple & $102$     & $139$   & $18$  & $49$ \\
 & Synthetic only    & single   & $1{,}408$ & $0$     & $306$ & $0$ \\
 & Synthetic only    & multiple & $1{,}788$ & $1{,}731$ & $320$ & $505$ \\
\cline{2-7}
 & \multicolumn{2}{l}{Subtotal} & $3{,}462$ & $1{,}878$ & $686$ & $557$ \\
\hline
\multicolumn{3}{l}{Total} & $4{,}325$ & $2{,}138$ & $1{,}527$ & $791$ \\
\hline\hline
\end{tabular}
\vspace{-0.6em}
\end{minipage}
\end{center}
\vspace{1.4em}

Table~\ref{tab:downselection_family_streamlined} breaks down the retained and dropped materials by structural family. The SuperCon and 3DSC$_{\mathrm{ICSD}}$ databases oversample the cuprates, most of whose compositions lack an exact ICSD match and carry only synthetically doped CIFs. The down selection therefore drops a disproportionately large share of the cuprates.

\begin{center}
\begin{minipage}{0.95\textwidth}
\centering
\refstepcounter{table}\label{tab:downselection_family_streamlined}
\parbox{\textwidth}{\textbf{TABLE~\thetable.} Structural-family composition of the retained and dropped superconducting (SC) and non-superconducting (non-SC) candidates. Family labels are assigned from composition and structure; ``alloy / intermetallic'' collects conventional superconductors not belonging to a named family.}
\vspace{-0.6em}
\begin{tabular}{l@{\hspace{0.8em}}r@{\hspace{0.8em}}r@{\hspace{0.8em}}r@{\hspace{0.8em}}r}
\hline\hline
 & \multicolumn{2}{c}{SC} & \multicolumn{2}{c}{non-SC} \\
\cline{2-5}
Family & kept & dropped & kept & dropped \\
\hline
Alloy / intermetallic & $1{,}859$ & $392$ & $747$ & $166$ \\
Cuprate               & $944$     & $1{,}173$ & $332$ & $420$ \\
Iron-based            & $401$     & $304$ & $88$ & $74$ \\
A15                   & $362$     & $40$ & $34$ & $7$ \\
Heavy-fermion         & $248$     & $9$ & $141$ & $15$ \\
Chalcogenide          & $216$     & $82$ & $77$ & $32$ \\
Oxide (other)         & $132$     & $110$ & $86$ & $68$ \\
MgB$_2$/AlB$_2$-type  & $130$     & $17$ & $8$ & $6$ \\
Chevrel               & $33$      & $11$ & $14$ & $3$ \\
\hline
Total                 & $4{,}325$ & $2{,}138$ & $1{,}527$ & $791$ \\
\hline\hline
\end{tabular}
\vspace{-0.6em}
\end{minipage}
\end{center}
\vspace{0.5em}

This curation step ensures that learned trends reflect intrinsic material properties rather than artifacts of structural ambiguity. With the featurized database and ground-truth $T_c$ values in hand, we train machine-learning models for two supervised tasks: $T_c$ prediction (regression) and superconductivity prediction (classification). The dataset is randomly split into 80\% training and 20\% test sets; model parameters are optimized on the training set and performance is evaluated on the held-out test set.

\subsection{Artificial doping procedure used in 3DSC}

Most chemical formulas in the SuperCon database are non-stoichiometric and therefore lack exact counterparts in the ICSD. To increase the number of matched entries, the 3DSC authors consider not only exact formula matches but also ICSD entries whose formulas are sufficiently similar to those in SuperCon, as quantified by several stoichiometry-based metrics detailed in Ref.~\cite{Sommer2023SciData}. When two formulas are deemed similar but not identical, their artificial-doping algorithm adjusts the site occupancies of the ICSD CIF to match the target composition. The resulting hypothetical CIF is then recorded together with the formula and $T_c$ from the SuperCon database.

Artificial doping starts from an ICSD crystal structure whose chemical formula is similar to the target SuperCon composition; this structure serves as a proxy for the actual, unmeasured structure. The algorithm partially replaces the atoms at specified crystallographic sites with the required dopant elements. Vacancies are handled by assigning zero scattering weight to the corresponding site fraction, effectively reducing the site occupancy. Only site occupancies are modified—atomic coordinates and interatomic distances remain unchanged. The underlying assumption is that formulas deemed similar by the 3DSC matching criteria correspond to structures with similar crystallographic parameters.

Three additional requirements constrain the artificial-doping procedure.
(a)~Each dopant must map to a unique set of equivalent crystallographic sites (i.e., sites sharing the same Wyckoff position). This mapping is unambiguous when the host element occupies the relevant sites in one of three ways: (i)~it fully occupies exactly one Wyckoff set and does not partially occupy any other; (ii)~it partially occupies exactly one set, while any additional fully occupied sets are disregarded; or (iii)~it partially occupies more than one set, but with identical occupancies on each, so that the sets can be treated collectively.
(b)~No crystallographic site may contain more than two elements after doping.
(c)~Doping must not add or remove crystallographic sites.
These constraints still permit application to complex compounds, including a large number of cuprates. The implementation is available at \url{https://github.com/aimat-lab/3DSC}.

\subsection{Graphlet feature generation from atomic properties}

In this work, we introduce a hierarchical graphlet expansion for material featurization. Starting from CIFs and a set of selected elemental properties, this framework encodes both chemical and structural information in a systematic manner.

The concept of graphlet expansion originated in biological network analysis, where entities such as proteins or genes are represented as identical nodes and graphlets are defined as small connected subgraphs. The frequency of these subgraphs serves as a key descriptor for characterizing complex networks. In those applications, only the topology of the graphlets is typically considered, while the specific properties of the nodes are discarded. In material featurization, however, we are interested not only in local connectivity but also in the chemical properties and geometry of the constituent atoms. To adapt graphlets to this context, we make two key enhancements. First, we retain the chemical identities of atoms when constructing the graphlets. Second, we use graphlets as the basis for encoding local chemical and structural properties, rather than simply recording graphlet frequencies. The procedure is described step by step below (see GitHub for details).

Structural representations of non-stoichiometric materials are necessarily approximate. Nevertheless, a significant fraction (65$\%$) of superconducting materials in our dataset are doped compounds. When one element partially substitutes another on a given site, we treat that site as occupied by an ``average'' atom whose elemental properties are occupancy-weighted averages of the constituent species. This virtual-atom approximation neglects changes in interatomic distances upon doping and treats the spatial distribution of dopants only in a mean-field sense. Despite these simplifications, the resulting features prove sufficient for our trained models to achieve $R_{\rm opt}^2>0.93$, as shown below.

\emph{Step 1:} We begin by reading the CIF of a material and identifying its primitive unit cell~\cite{ONG2013Commatsci}. For each atomic site in the cell, we record the constituent elements and their occupancies (relevant for doped or vacancy-bearing sites) and identify nearest neighbors using the \texttt{VoronoiNN} algorithm, which includes neighbors in adjacent periodic images. A neighbor is retained as valid if it lies within a cutoff distance of 1.5 times the sum of the atomic radii of the two sites, using the empirical radii published by Slater~\cite{atomic_radii}. For a partially occupied site, the effective atomic radius is the occupancy-weighted average of the constituent species' radii. For example, in Mg$_{0.95}$Al$_{0.05}$B$_2$, the B atoms fully occupy the 2d sites with an effective radius of 85~pm, while the 1a site yields an effective radius of $(0.95 \times 150~\text{pm}) + (0.05 \times 125~\text{pm}) = 148.75~\text{pm}$.

\begin{figure}
    \centering
    \includegraphics[width=0.5\linewidth]{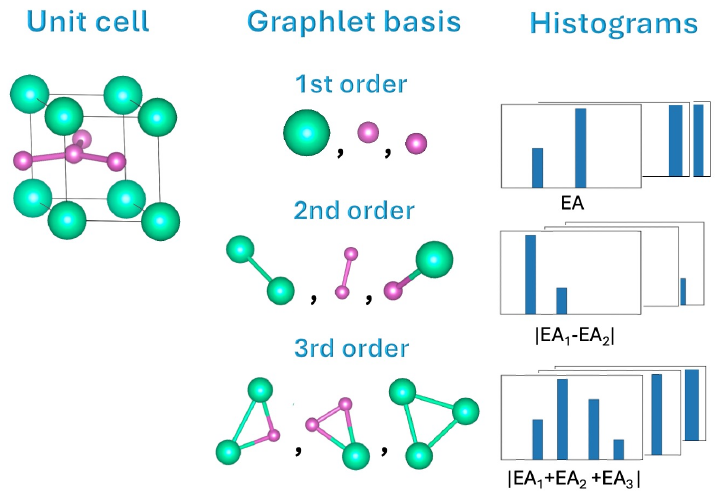}
    \caption{Illustration of the complete graphlet basis up to third order and the corresponding histogram features, generated from a primitive unit cell. Graphlets are defined in terms of crystallographic sites, not elements: two distinct sites occupied by the same element (purple) constitute inequivalent first-order graphlets. Histogram features are constructed from permutation-invariant combinations of the site-level chemical features.}
    \label{fig:feature}
\end{figure}

\emph{Step 2:} We enumerate all atomic sites and their valid neighbors to construct the complete sets of first-, second-, and third-order graphlets. A first-order graphlet is a single crystallographic site; the complete first-order set comprises all inequivalent sites in the primitive cell. A second-order graphlet is a valid nearest-neighbor pair; its complete set includes all inequivalent such pairs. A third-order graphlet is a center site together with two of its valid neighbors (which need not be valid neighbors of each other); its complete set collects all inequivalent such triangles. These three graphlet orders are illustrated in Fig.~\ref{fig:feature}.

\begin{table}[ht!]
\centering
\caption{List of elemental properties used for constructing chemical features.}
\begin{tabular}{llp{0.3\linewidth}p{0.3\linewidth}}
\hline
\textbf{Property} & \textbf{Description} & \raggedright \textbf{Source} & \raggedright \textbf{Method} \tabularnewline
\hline
EN     & Pauling electronegativity & \raggedright CRC Handbook of Chemistry and Physics~\cite{Rumble2020} & \raggedright Based on measurement (semi-empirical) \tabularnewline
EA     & Electron affinity (eV)     & \raggedright CRC Handbook of Chemistry and Physics & \raggedright Measured; few elements are calculated \tabularnewline
IP     & Ionization potential (eV)  & \raggedright NIST Atomic Spectra Database Ionization Energies Form~\cite{NIST_IonEnergy_2011} & \raggedright Measured; few elements are calculated \tabularnewline
R\textsubscript{cov} & Covalent radius (pm) & \raggedright CRC Handbook of Chemistry and Physics & \raggedright Based on measurement; few elements are calculated \tabularnewline
AW     & Atomic weight              & \raggedright Atomic weights of the elements 2013~\cite{Meija2016} & \raggedright Based on measurement \tabularnewline
N\textsubscript{s}   & Valence electrons in $s$ & \raggedright Periodic Table & \raggedright Determined from the periodic table \tabularnewline
N\textsubscript{p}   & Valence electrons in $p$ & \raggedright Periodic Table & \raggedright Determined from the periodic table \tabularnewline
N\textsubscript{d}   & Valence electrons in $d$ & \raggedright Periodic Table & \raggedright Determined from the periodic table \tabularnewline
N\textsubscript{tot} & Total valence electrons  & \raggedright Periodic Table & \raggedright Determined from the periodic table \tabularnewline
Col    & Column number in periodic table & \raggedright Periodic Table & \raggedright Determined from the periodic table \tabularnewline
\hline
\end{tabular}
\label{tab:elemental_properties}
\end{table}

\emph{Step 3:} We assign chemical and structural features to each graphlet constructed in Step~2. The chemical features are built from the 10 elemental properties listed in Table~\ref{tab:elemental_properties}. For a first-order graphlet (a single site), each elemental feature is the occupancy-weighted average over all constituent species. For a second-order graphlet (a nearest-neighbor pair), we assign both chemical and structural features. The structural feature is the intersite distance. The chemical features must be permutation-invariant with respect to the two sites; for each elemental property, we therefore take the mean and the absolute difference of the two site values, yielding $2 \times 10 = 20$ chemical features plus one distance, for a total of 21 features per pair. For a third-order graphlet (a triangle of three sites), the structural degrees of freedom are three pairwise distances and three angles. We again require permutation invariance: for both chemical and structural quantities, we take the mean, standard deviation, and kurtosis over the three sites or pairs, yielding $3 \times 10 = 30$ chemical features and $3 \times 2 = 6$ structural features, for a total of 36. In summary, the graphlet expansion of a material comprises 10 first-order, 21 second-order, and 36 third-order features.

\emph{Step 4:} We convert the graphlet expansion of each material into fixed-length histogram features. Machine-learning models typically require inputs of uniform size, yet different crystal structures yield different numbers of graphlets. Because all materials share the same three graphlet orders and the same feature set within each order, we can represent each feature as a distribution over the graphlets of a given order. Each such distribution is discretized into a histogram $h=\{(m^{(1)},\;h^{(1)}),\;(m^{(2)},\;h^{(2)}),\;\ldots,\;(m^{(n)},\;h^{(n)}) \}$, where $h^{(i)}$ is the number of graphlet feature values that fall into the $i$th bin, centered at $m^{(i)}$ (see Fig.~\ref{fig:feature}). Each material is thus described by 10 first-order, 21 second-order, and 36 third-order histograms. All histograms use 20 bins, and for each feature the bin range is fixed across materials by the global minimum and maximum of that feature over the entire dataset.

\emph{Step 5:} We standardize the bin midpoints of each histogram feature to ensure comparability across features with different physical units and scales. For a given feature, let $\{m^{(i)}\}$ denote its bin midpoints and $\{c^{(i)}\}$ the corresponding counts aggregated over all materials in the dataset. The weighted mean $\bar{m}$ and standard deviation $\sigma$ are
\begin{equation}
\bar{m} = \frac{\sum_i m^{(i)} c^{(i)}}{\sum_i c^{(i)}}, \qquad
\sigma = \sqrt{ \frac{\sum_i m^{(i)2} c^{(i)}}{\sum_i c^{(i)}} - \left( \bar{m} \right)^2 }.
\end{equation}
Each bin midpoint $m^{(i)}$ is then standardized as
\begin{equation}
m^{(i)}_s = \frac{m^{(i)} - \bar{m}}{\sigma},
\end{equation}
while the bin counts $c^{(i)}$ remain unchanged. This standardization is applied independently to each of the 67 histogram features, producing midpoints that are zero-centered and unit-scaled. The normalization is essential for the additive EMD kernel used in our Gaussian process model (described below), which is sensitive to the scale of the feature space.

\subsection{Symmetry features}

Crystallographic symmetry plays an essential role in determining material properties, yet its treatment in machine learning for condensed matter physics remains at an early stage. On the feature level, some previous ML studies encode symmetry simply by supplying the space group number as an input, but this is a human-defined label that carries no intrinsic physical meaning for the model. On the modeling side, there has been growing interest in equivariant neural networks~\cite{batzner2022communications,geiger2022e3nn}, which ensure that model outputs transform appropriately under symmetry operations applied to the inputs. While such networks faithfully respect all symmetry operations of a given group (e.g., the Euclidean group), they are inherently agnostic to which specific symmetry elements are relevant to the target property and thus cannot learn to distinguish symmetry–property relationships.

In this work, we approach the problem from the perspective of feature design and introduce a principled method for encoding crystal symmetry into physically meaningful features. The symmetry of a crystalline material is described by its space group. Since space groups are infinite (due to lattice translations), directly converting them into finite-dimensional features is challenging. Space group operations can be represented as Seitz pairs $(\mathbf{W}, \mathbf{w})$, where a point $\mathbf{x}$ is mapped to $\mathbf{W}\mathbf{x} + \mathbf{w}$. The set of all rotational parts $\mathbf{W}$ of a space group $\mathcal{G}$ forms its point group $\mathcal{P}$~\cite{InternationalTables2016}. In three dimensions there are only 32 crystallographic point groups, all finite. A natural first approach would be to construct symmetry indicators at the point-group level, but this coarse-graining incurs significant information loss: many distinct space groups share the same point group and would receive identical indicators.

To recover finer-grained symmetry information, we observe that materials sharing the same space group may differ in which Wyckoff positions are occupied. We therefore assign symmetry indicators at the level of individual sites rather than the full space group. The subgroup $S_{\mathbf{x}} \subset \mathcal{G}$ of symmetry operations that leave a crystallographic site $\mathbf{x}$ invariant is its site-symmetry group~\cite{InternationalTables2016}. Every site-symmetry group in three dimensions is isomorphic to one of the 32 crystallographic point groups, since it is a subgroup of the material's point group $\mathcal{P}$. We can therefore design symmetry indicators for each of the 32 point groups and apply them to the site-symmetry group of every occupied site.

\begin{figure}[ht!]
    \centering
    \includegraphics[width=\linewidth]{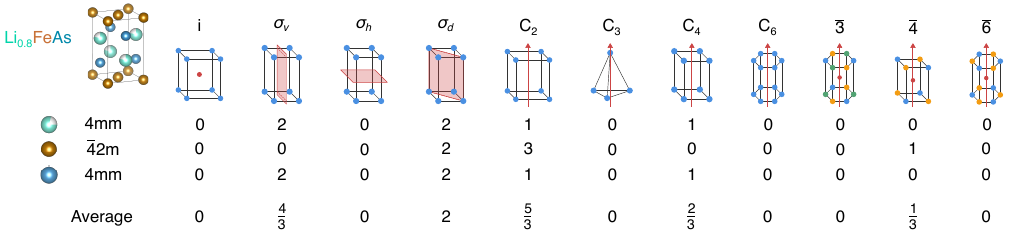}
    \caption{
    The eleven point-group operation types assigned to each site according to its site-symmetry group.
    The crystal symmetry vector is the average of the site-symmetry vectors over all inequivalent occupied sites in the unit cell.
    }
    \label{fig:Symmetry_LiFeAs}
\end{figure}

In the Schoenflies notation, there are 11 distinct types of point-group symmetry operations (excluding the identity), summarized in Table~\ref{tab:pg_operations}. We accordingly define the symmetry indicator as an 11-dimensional vector, with each component corresponding to one operation type. For a given point group $\mathcal{P}$, each component records the number of distinct geometric elements—inversion centers, mirror planes, or rotation axes—that generate the corresponding operation. We count geometric elements rather than operations because a single geometric element can generate multiple operations (e.g., a $C_4$ axis generates three distinct rotations, whereas a $C_2$ axis generates only one); counting operations directly would weight higher-order axes disproportionately. Table~\ref{tab:pg_symmetry_vectors} lists the resulting 11-dimensional symmetry vectors for all 32 crystallographic point groups (see also Fig.~\ref{fig:Symmetry_LiFeAs}).

\begin{table}[ht]
\centering
\caption{The 11 types of point group symmetry operations (excluding identity), following the Schoenflies notation.}
\label{tab:pg_operations}
\begin{tabular}{lll}
\hline
\textbf{Symbol} & \textbf{Name} & \textbf{Description} \\
\hline
$i$            & Inversion                    & Inversion through the origin: $\vec{r} \rightarrow -\vec{r}$ \\
$\sigma_v$     & Vertical mirror plane        & Mirror plane parallel to the principal rotation axis \\
$\sigma_h$     & Horizontal mirror plane      & Mirror plane perpendicular to the principal rotation axis \\
$\sigma_d$     & Dihedral mirror plane        & Mirror plane at diagonal angles between vertical planes \\
$C_2$          & Two-fold rotation            & Rotation by $180^\circ$ about the principal axis \\
$C_3$          & Three-fold rotation          & Rotation by $120^\circ$ about the principal axis \\
$C_4$          & Four-fold rotation           & Rotation by $90^\circ$ about the principal axis \\
$C_6$          & Six-fold rotation            & Rotation by $60^\circ$ about the principal axis \\
$\bar{3}$          & Three-fold improper rotation & $C_3$ rotation followed by reflection through perpendicular plane \\
$\bar{4}$          & Four-fold improper rotation  & $C_4$ rotation followed by reflection through perpendicular plane \\
$\bar{6}$          & Six-fold improper rotation   & $C_6$ rotation followed by reflection through perpendicular plane \\
\hline
\end{tabular}
\end{table}

\begin{figure}[ht!]
    \centering
    \includegraphics[width=0.7\linewidth]{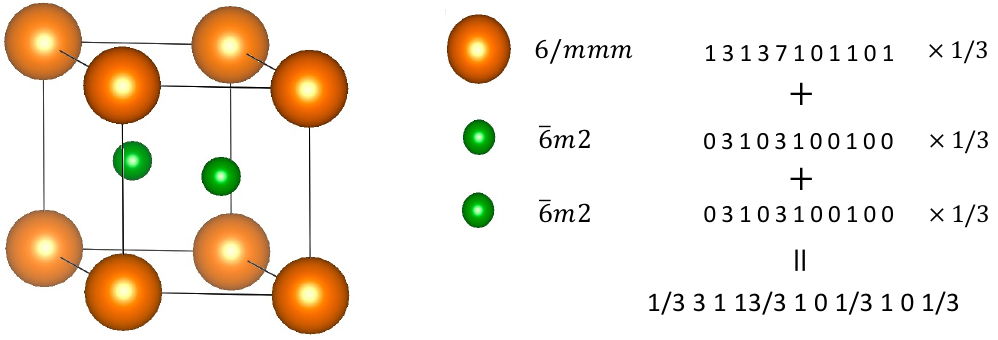}
    \caption{Symmetry feature construction for MgB$_2$. The site-symmetry group of the Mg site is isomorphic to point group $6/mmm$, while that of the two B sites is isomorphic to $\bar{6}m2$. The corresponding 11-dimensional symmetry indicators are listed in Table~\ref{tab:pg_symmetry_vectors}. The symmetry feature of the material is the component-wise average of the three site-symmetry vectors.
    }
    \label{fig:Symmetry_MgB2}
\end{figure}

With the 32 point-group indicators in hand, we generate the symmetry feature for each material as follows. Our implementation uses two Python libraries, \texttt{pymatgen} and \texttt{spglib}. First, we identify all inequivalent atomic sites in the unit cell. Using \texttt{SpaceGroupAnalyzer.get\_symmetry\_operations()} from \texttt{pymatgen}, we obtain the set of symmetry operations sufficient for determining site-symmetry groups. For each occupied site $\mathbf{x}$, we retain the operations $(\mathbf{W}, \mathbf{w})$ satisfying $\mathbf{W}\mathbf{x} + \mathbf{w} \equiv \mathbf{x}$, i.e., those that leave $\mathbf{x}$ invariant. The rotational parts $\mathbf{W}$ of these operations are passed to \texttt{spglib.get\_pointgroup()}, which returns the corresponding crystallographic point group. Once the point group of every inequivalent site has been identified, its 11-dimensional symmetry indicator is retrieved, and the final symmetry feature vector for the material is the component-wise average over all inequivalent sites. The procedure is illustrated in Fig.~\ref{fig:Symmetry_LiFeAs} for Li$_{0.8}$FeAs and in Fig.~\ref{fig:Symmetry_MgB2} for MgB$_2$.

\begin{table}[ht]
\centering
\caption{Symmetry feature vectors for the 32 crystallographic point groups. Each row corresponds to one point group and is essentially an 11-dimensional symmetry vector, with each column indicating the number of associated geometric elements for the corresponding symmetry operation type.}
\label{tab:pg_symmetry_vectors}
\footnotesize
\setlength{\tabcolsep}{9pt} 
\begin{tabular}{lccccccccccc}
\hline
\textbf{Point Group} & $i$ & $\sigma_v$ & $\sigma_h$ & $\sigma_d$ & $C_2$ & $C_3$ & $C_4$ & $C_6$ & $\bar{3}$ & $\bar{4}$ & $\bar{6}$ \\
\hline
$1$           & 0 & 0 & 0 & 0 & 0 & 0 & 0 & 0 & 0 & 0 & 0 \\
$\bar{1}$     & 1 & 0 & 0 & 0 & 0 & 0 & 0 & 0 & 0 & 0 & 0 \\
$2$           & 0 & 0 & 0 & 0 & 1 & 0 & 0 & 0 & 0 & 0 & 0 \\
$m$           & 0 & 0 & 1 & 0 & 0 & 0 & 0 & 0 & 0 & 0 & 0 \\
$2/m$         & 1 & 0 & 1 & 0 & 1 & 0 & 0 & 0 & 0 & 0 & 0 \\
$222$         & 0 & 0 & 0 & 0 & 3 & 0 & 0 & 0 & 0 & 0 & 0 \\
$mm2$         & 0 & 2 & 0 & 0 & 1 & 0 & 0 & 0 & 0 & 0 & 0 \\
$mmm$         & 1 & 2 & 1 & 0 & 3 & 0 & 0 & 0 & 0 & 0 & 0 \\
$4$           & 0 & 0 & 0 & 0 & 1 & 0 & 1 & 0 & 0 & 0 & 0 \\
$\bar{4}$     & 0 & 0 & 0 & 0 & 1 & 0 & 0 & 0 & 0 & 1 & 0 \\
$4/m$         & 1 & 0 & 1 & 0 & 1 & 0 & 1 & 0 & 0 & 1 & 0 \\
$422$         & 0 & 0 & 0 & 0 & 5 & 0 & 1 & 0 & 0 & 0 & 0 \\
$4mm$         & 0 & 2 & 0 & 2 & 1 & 0 & 1 & 0 & 0 & 0 & 0 \\
$\bar{4}2m$   & 0 & 0 & 0 & 2 & 3 & 0 & 0 & 0 & 0 & 1 & 0 \\
$4/mmm$       & 1 & 2 & 1 & 2 & 5 & 0 & 1 & 0 & 0 & 1 & 0 \\
$3$           & 0 & 0 & 0 & 0 & 0 & 1 & 0 & 0 & 0 & 0 & 0 \\
$\bar{3}$     & 1 & 0 & 0 & 0 & 0 & 1 & 0 & 0 & 0 & 0 & 1 \\
$32$          & 0 & 0 & 0 & 0 & 3 & 1 & 0 & 0 & 0 & 0 & 0 \\
$3m$          & 0 & 3 & 0 & 0 & 0 & 1 & 0 & 0 & 0 & 0 & 0 \\
$\bar{3}m$    & 1 & 0 & 0 & 3 & 3 & 1 & 0 & 0 & 0 & 0 & 1 \\
$6$           & 0 & 0 & 0 & 0 & 1 & 1 & 0 & 1 & 0 & 0 & 0 \\
$\bar{6}$     & 0 & 0 & 1 & 0 & 0 & 1 & 0 & 0 & 1 & 0 & 0 \\
$6/m$         & 1 & 0 & 1 & 0 & 1 & 1 & 0 & 1 & 1 & 0 & 1 \\
$622$         & 0 & 0 & 0 & 0 & 7 & 1 & 0 & 1 & 0 & 0 & 0 \\
$6mm$         & 0 & 3 & 0 & 3 & 1 & 1 & 0 & 1 & 0 & 0 & 0 \\
$\bar{6}m2$   & 0 & 3 & 1 & 0 & 3 & 1 & 0 & 0 & 1 & 0 & 0 \\
$6/mmm$       & 1 & 3 & 1 & 3 & 7 & 1 & 0 & 1 & 1 & 0 & 1 \\
$23$          & 0 & 0 & 0 & 0 & 3 & 4 & 0 & 0 & 0 & 0 & 0 \\
$m\bar{3}$    & 1 & 0 & 3 & 0 & 3 & 4 & 0 & 0 & 0 & 0 & 4 \\
$432$         & 0 & 0 & 0 & 0 & 9 & 4 & 3 & 0 & 0 & 0 & 0 \\
$\bar{4}3m$   & 0 & 0 & 0 & 6 & 3 & 4 & 0 & 0 & 0 & 3 & 0 \\
$m\bar{3}m$   & 1 & 0 & 3 & 6 & 9 & 4 & 3 & 0 & 0 & 3 & 4 \\
\hline
\end{tabular}
\end{table}

\newpage

\begin{figure}[ht!]
    \centering
    \includegraphics[width=0.9\linewidth]{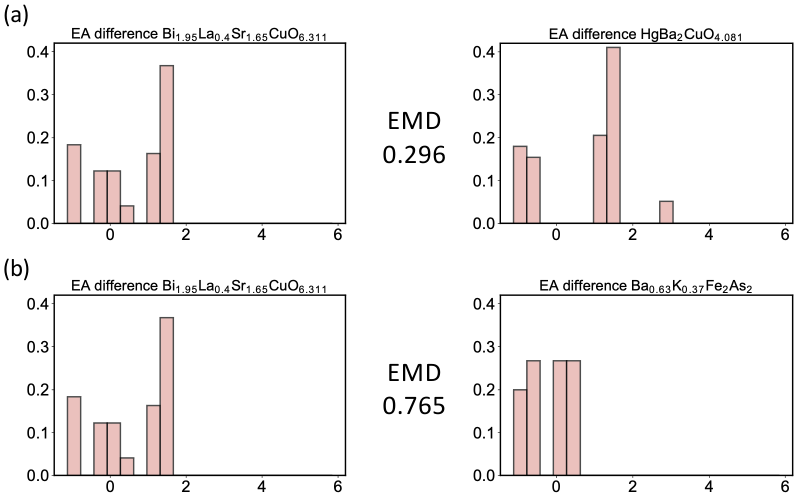}
    \caption{Examples of EMD computation between pairs of materials. (a)~EMD between second-order electron-affinity-difference histograms of two cuprate materials: the distributions are similar and the EMD is small. (b)~The same comparison between a cuprate and an iron-based material: the distributions differ substantially and the EMD is large.}
    \label{fig:EMD_example}
\end{figure}

\section{Gaussian process based on the EMD kernel}
\begin{figure}
    \centering
    \includegraphics[width=\linewidth]{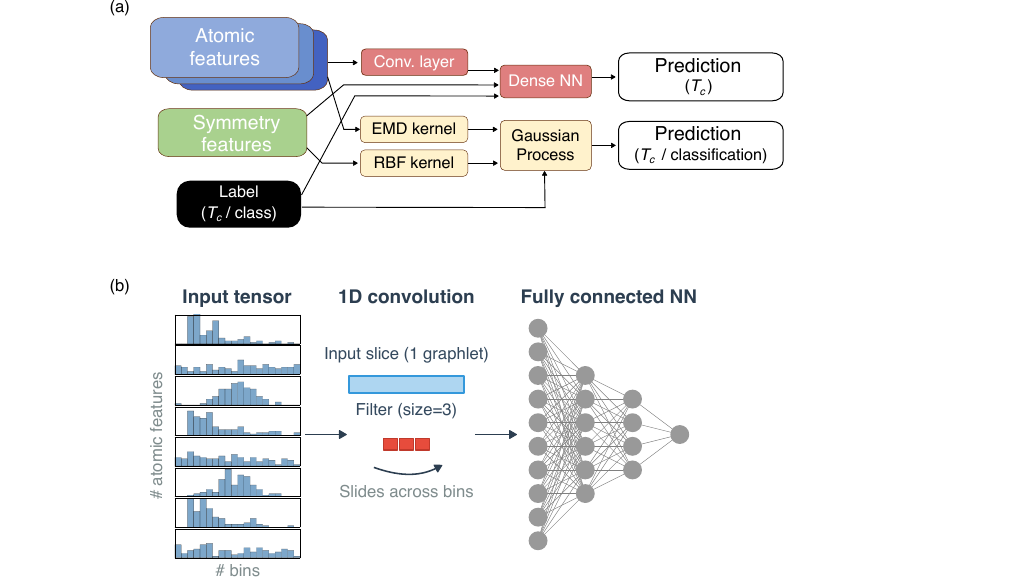}
    \caption{
    (a)~Flowchart of the machine learning workflow. Graphlet histograms and symmetry vectors are input to neural networks and GP regressors for $T_c$ prediction, and to a GP classifier for superconductivity classification.
    (b)~Architecture of the neural network baseline: the graphlet histograms are processed by a 1D convolutional layer before being flattened and passed to a fully connected feed-forward network.
    }
    \label{fig:ML-workflow}
\end{figure}

For $T_c$ prediction, we compare a flexible but opaque neural network (NN) baseline against a transparent, probabilistic Gaussian process (GP) model, as illustrated in Fig.~\figref{fig:ML-workflow}{a}. We deploy GP models for two distinct tasks: (i)~regression to predict the critical temperature $T_c$, and (ii)~binary classification to predict whether a material is a superconductor. For both tasks, the input features are either the graphlet histograms alone or the histograms augmented with the symmetry features. Graphlet features are compared using the Earth Mover's Distance (EMD), from which we construct an additive EMD kernel. Symmetry features are modeled using a standard squared-exponential Automatic Relevance Determination (ARD) kernel. In the following, we define the EMD and EMD-based kernel, prove its validity, and then detail the GP regression and classification frameworks.

Symmetry features, which are finite-dimensional vectors, are modeled using the squared-exponential ARD kernel, and the joint covariance over both graphlet and symmetry features is constructed as a product of the EMD and ARD kernels. We note that the inputs to these kernels are standardized differently: the bin centers of the graphlet histograms are normalized via the procedure in Step 5, whereas the symmetry feature values are scaled to the interval $[0,1]$. This GP formulation with an EMD kernel provides a principled framework for learning from distribution-valued features, preserving their full shape information while allowing for feature sensitivity analysis.

\subsection{Earth mover's distance}
The Earth Mover's Distance (EMD)~\cite{rubner_earth_2000} is a metric that quantifies the dissimilarity between two distributions. It has found widespread applications in computer vision and physics, such as characterizing jet substructure in particle collider experiments~\cite{komiske_metric_2019}.
Mathematically, EMD compares two signatures $A=\{(\mathbf{a}_1, w_{\mathbf{a}_1}), \ldots, (\mathbf{a}_n, w_{\mathbf{a}_n})\}$ and $B=\{(\mathbf{b}_1, w_{\mathbf{b}_1}), \ldots, (\mathbf{b}_m, w_{\mathbf{b}_m})\}$, where $\mathbf{a}_i$ ($\mathbf{b}_j$) represents the coordinate of a cluster or bin center, and $w_{\mathbf{a}_i}$ ($w_{\mathbf{b}_j}$) is its corresponding weight. In the general case where the total weights of $A$ and $B$ are not normalized and the number of clusters differ ($n \neq m$), we solve for a transport flow matrix $F=[f_{ij}]$—where $f_{ij} \geq 0$ is the mass transported from $\mathbf{a}_i$ to $\mathbf{b}_j$—that minimizes the work
\begin{equation}
    \mathrm{work}(A,B,F)=\sum_{ij}d_{ij}f_{ij},
\end{equation}
subject to the constraints:
\begin{align}
    f_{ij} &\geq 0, \quad 1\leq i\leq n, \; 1\leq j\leq m \\
    \sum_{j=1}^m f_{ij} &\leq w_{\mathbf{a}_i}, \quad 1\leq i\leq n \\
    \sum_{i=1}^n f_{ij} &\leq w_{\mathbf{b}_j}, \quad 1\leq j\leq m \\
    \sum_{i=1}^n \sum_{j=1}^m f_{ij} &= \min\left(\sum_{i=1}^n w_{\mathbf{a}_i}, \; \sum_{j=1}^m w_{\mathbf{b}_j}\right),
\end{align}
where $d_{ij}=\lVert\mathbf{a}_i-\mathbf{b}_j\rVert$. Intuitively, representing $A$ as piles of sand and $B$ as holes, EMD measures the minimum cost of moving the sand to fill the holes, maximizing total flow up to the smaller of the two masses. The EMD is then the minimized work normalized by the total flow:
\begin{equation}
    \mathrm{EMD}(A,B) = \frac{\sum_{i=1}^n \sum_{j=1}^m d_{ij} f_{ij}}{\sum_{i=1}^n \sum_{j=1}^m f_{ij}}.
\end{equation}

Our graphlet features are discrete histograms $h = \{(m^{(1)}, h^{(1)}), (m^{(2)}, h^{(2)}), \ldots, (m^{(n)}, h^{(n)})\}$, where each feature dimension shares equally spaced bin centers $\{m^{(i)}\}$. To eliminate the normalization denominator, we scale each histogram such that $\sum_{i=1}^n h^{(i)} = 1$. The EMD between two normalized 1D histograms then simplifies to
\begin{equation}\label{eq:EMD_def}
    \text{EMD}(h_1, h_2) = \min_{F} \sum_{i=1}^{n} \sum_{j=1}^{n} \Delta m \lvert i-j \rvert f_{ij},
\end{equation}
where $\Delta m = m^{(i)} - m^{(i-1)}$ is the constant bin width. Representative pairs of histogram features with small and large EMD values are shown in Fig.~\ref{fig:EMD_example}.

Given that the EMD defines a physically meaningful distance between individual histograms, we construct an additive EMD kernel to compare two feature vectors of histograms, $\mathbf{x}_i = (h_{i1}, h_{i2}, \ldots, h_{iq})$ and $\mathbf{x}_j = (h_{j1}, h_{j2}, \ldots, h_{jq})$, where $q$ is the number of histogram features (in our work, $q=21$ for the second-order graphlet features, and $q=57$ when the 36 third-order features are additionally included). The additive EMD kernel is defined as
\begin{equation}
    k_{\mathrm{EMD}}(\mathbf{x}_i, \mathbf{x}_j) = \sum_{n=1}^q w_n \exp\left(-\frac{\mathrm{EMD}(h_{in}, h_{jn})}{\ell_n}\right),
\end{equation}
where $\{w_n\}$ and $\{\ell_n\}$ are learnable parameters. Because histogram distributions can have complex shapes and a few dimensions might have highly dissimilar features (yielding large EMD values), the additive form aggregates per-dimension similarities to ensure that strong similarity along informative features is not suppressed. This prevents the kernel from collapsing to small values, as would occur with a multiplicative product dominated by a few "bad" dimensions. In our implementation, individual 1D EMD values are computed using the \texttt{wasserstein\_1d} function in the Python Optimal Transport (POT) library~\cite{POT}.

When using only graphlet features, the GP models directly employ the additive EMD kernel. When incorporating both graphlet and symmetry features, we employ a multiplicative product of the EMD kernel and the symmetry ARD kernel: $k(\mathbf{x}_i, \mathbf{x}_j) = \sigma^2 k_{\mathrm{EMD}}(\mathbf{x}_i, \mathbf{x}_j) \cdot k_{\mathrm{ARD}}(\mathbf{s}_i, \mathbf{s}_j)$, where $\mathbf{s}$ represents the symmetry vector. This coupling yields high covariance only when inputs are similar in both feature spaces, encouraging the GP model to leverage joint information.

\subsection{Proof of EMD kernel validity}
\label{sec:kernel_proof}

For successful GP-based learning with our unique histogram features, we need to construct a suitable and valid kernel based on a meaningful metric in the feature space. The EMD already used to navigate the multiplicity of CIFs will be a natural metric. However, constructing a valid Mercer kernel from the EMD is nontrivial. In more than one dimension the EMD is nonlinear, and the transformations routinely applied to Euclidean distances in machine learning (such as $\psi(r):=\exp(-r^2)$) do not necessarily apply, because the EMD Gram matrix is not conditionally negative definite. For instance, while it is tempting to construct a kernel that mimics more popular kernels, such as the radial basis function (RBF) kernel, e.g., by squaring the EMD, this does not result in a valid kernel.

As defined in Eq.~\eqref{eq:EMD_def}, the EMD with an L1 ground metric is given by
\begin{equation*}
    \text{EMD}(h_1, h_2) = \min_{F} \left( d\sum_{i,j} |i-j| f_{ij} \right).
\end{equation*}
Let $C(h) \in \mathbb{R}^{n}$ be the cumulative histogram of $h$:
\begin{equation*}
    [C(h)]_{k} \triangleq \sum_{b=1}^{k}h_{b},
\end{equation*}
where $[C(h)]_{k}$ denotes the $k$th bin of the cumulative histogram. A well-known property of the 1D EMD with L1 ground distance is a reduction to an L1 distance (see e.g. Ref.~\cite{martinez2016closed}):
\begin{equation}\label{eq:emd_cdf}
    \textrm{EMD}(h_1, h_2) = d \Vert C(h_1) - C(h_2) \Vert_{1},
\end{equation}
where again $d$ is the distance between bin centers. Because the L1 distance $\Vert u - v\Vert_1$ is conditionally negative definite, by Schoenberg's theorem, the function
\begin{equation*}
    k_{\lambda}(u, v) = \exp(-\lambda\Vert u - v \Vert_1)
\end{equation*}
is positive definite for any $\lambda > 0$. Choosing $\lambda = d/\ell_\alpha$, which is positive for any length scale $\ell_\alpha > 0$, and using Eq.~\eqref{eq:emd_cdf}, we obtain
\begin{equation}
    k(h_1, h_2) := \exp\left(-\frac{d}{\ell_\alpha} \Vert C(h_1) - C(h_2)\Vert_1\right) = \exp\left(-\frac{{\rm EMD}(h_1, h_2)}{\ell_\alpha}\right)
\end{equation}
is positive definite (PD). This demonstrates that the kernel restricted to a single feature dimension is valid. Finally, non-negative weighted sums of PD kernels are PD (since $v^{\top}\left[\sum w_{i}K_{i}\right]v = \sum w_{i}v^{\top}K_{i}v > 0$). Therefore, the final kernel expression,
\begin{equation}
    k_{\rm EMD}(x_i, x_j) = \sum_{\alpha} w_{\alpha} \exp\left(-\frac{\mathrm{EMD}(h_{i,\alpha}, h_{j,\alpha})}{\ell_{\alpha}}\right),
\end{equation}
is positive definite because it is a positively weighted sum of the ``individual histogram EMD'' kernel that we just proved is a valid kernel.

\subsection{Details of the Gaussian process models}

Having constructed a valid kernel for the histogram features, we frame the regression and classification models. Given training data $\mathcal{D}=\{(\mathbf{x}_i, y_i)\}_{i=1}^N$, where $\mathbf{x}_i$ contains either the graphlet features alone or the augmented graphlet and symmetry features, and the targets are the critical temperatures $y_i = T_{\text{c}}^{(i)}$, we place a Gaussian process prior on the latent function $f$:
\begin{equation}\label{eq:gp_prior}
f \sim \mathcal{GP}\!\big(m(\cdot),\,k(\cdot,\cdot;\theta)\big),
\end{equation}
where $m(\cdot)$ is the mean function and $k(\cdot,\cdot;\theta)$ is the covariance function parameterized by hyperparameters $\theta$. The kernel $k$ is either $k_{\mathrm{EMD}}$ or the product kernel $k = \sigma^2 k_{\mathrm{EMD}} \cdot k_{\mathrm{ARD}}$.

The regression observations are assumed to follow a Gaussian noise model:
\begin{equation}
y_i = f(\mathbf{x}_i) + \varepsilon_i,\qquad \varepsilon_i \sim \mathcal{N}(0,\sigma_n^2).
\end{equation}

Let $X=[\mathbf{x}_1, \ldots, \mathbf{x}_N]$, $\mathbf{y}=[y_1, \ldots, y_N]^\top$, $\mathbf{m}_X=[m(\mathbf{x}_1), \ldots, m(\mathbf{x}_N)]^\top$, and the Gram matrix $K_{XX}=[k(\mathbf{x}_i, \mathbf{x}_j)]_{i,j}$, with $K_y = K_{XX}+\sigma_n^2 I$. For a test input $\mathbf{x}_\star$, we define the covariance vector $\mathbf{k}_\star=[\,k(\mathbf{x}_1, \mathbf{x}_\star), \ldots, k(\mathbf{x}_N, \mathbf{x}_\star)\,]^\top$. The exact GP posterior over the latent function value $f(\mathbf{x}_\star)$ is Gaussian with mean $\mu_\star$ and variance $v_\star$:
\begin{equation}
\mu_\star = m(\mathbf{x}_\star) + \mathbf{k}_\star^\top K_y^{-1}\big(\mathbf{y}-\mathbf{m}_X\big),
\end{equation}
\begin{equation}
v_\star = k(\mathbf{x}_\star, \mathbf{x}_\star) - \mathbf{k}_\star^\top K_y^{-1} \mathbf{k}_\star.
\end{equation}
The predictive distribution for a noisy observation $y_\star$ is given by $\mathrm{Var}(y_\star) = v_\star + \sigma_n^2$.

The kernel hyperparameters $\theta$ (consisting of the EMD parameters $\{w_n, \ell_n\}_{n=1}^q$, the signal amplitude $\sigma^2$, the noise variance $\sigma_n^2$, and any parameters of the mean function) are optimized by maximizing the log marginal likelihood:
\begin{equation}
\log p(\mathbf{y}\,|\,X,\theta)
= -\tfrac12 (\mathbf{y}-\mathbf{m}_X)^\top K_y^{-1}(\mathbf{y}-\mathbf{m}_X)
  - \tfrac12 \log|K_y|
  - \tfrac{N}{2}\log(2\pi).
\end{equation}
Gradients of this objective with respect to $\theta$ are computed analytically to facilitate gradient-based optimization.

For the binary classification task ($y_i \in \{-1, +1\}$), the likelihood is non-Gaussian, making exact inference intractable. We employ a variational Gaussian process classifier. We assume the same GP prior on the latent function $f$ as in Eq.~\eqref{eq:gp_prior}. To obtain a tractable approximation, we introduce inducing inputs $\mathbf{Z}=[\mathbf{z}_1, \ldots, \mathbf{z}_M]$ and corresponding inducing variables $\mathbf{u}=f(\mathbf{Z})$ with prior
\begin{equation}
p(\mathbf{u}) = \mathcal{N}\!\big(\mathbf{m}_Z,\,K_{ZZ}\big),
\end{equation}
where $\mathbf{m}_Z = [\,m(\mathbf{z}_1), \ldots, m(\mathbf{z}_M)\,]^\top$ and $K_{ZZ}=[\,k(\mathbf{z}_i, \mathbf{z}_j)\,]_{i,j}$. In our setting, we set the inducing inputs equal to the training inputs, $\mathbf{Z}=\mathbf{X}$ (so $M=N$); thus, the inducing variables function as a variational device to ensure tractability rather than a sparse approximation. We approximate the intractable true posterior with a Gaussian variational posterior:
\begin{equation}
q(\mathbf{u}) = \mathcal{N}\!\big(\boldsymbol{\mu}_u,\,\mathbf{S}_u\big),
\end{equation}
which induces the approximate posterior process
\begin{equation}
q\big(f(\cdot)\big) = \int p\big(f(\cdot)\,\big|\,\mathbf{u}\big)\, q(\mathbf{u})\, d\mathbf{u}.
\end{equation}

For a test input $\mathbf{x}_\star$, let $\mathbf{k}_{Z\star}=[\,k(\mathbf{z}_1,\mathbf{x}_\star), \ldots, k(\mathbf{z}_M, \mathbf{x}_\star)\,]^\top$ and $k_{\star\star}=k(\mathbf{x}_\star, \mathbf{x}_\star)$. The resulting predictive marginal $q(f(\mathbf{x}_\star))$ is a univariate Gaussian $\mathcal{N}(\mu_\star, v_\star)$ with:
\begin{equation}
\mu_\star = m(\mathbf{x}_\star) + \mathbf{k}_{Z\star}^\top K_{ZZ}^{-1}\big(\boldsymbol{\mu}_u - \mathbf{m}_Z\big),
\end{equation}
\begin{equation}
v_\star = k_{\star\star} + \mathbf{k}_{Z\star}^\top K_{ZZ}^{-1}\big(\mathbf{S}_u - K_{ZZ}\big)K_{ZZ}^{-1}\mathbf{k}_{Z\star}.
\end{equation}

Under a probit likelihood model, we have
\begin{equation}
p\big(y_i \mid f(\mathbf{x}_i)\big) = \Phi\!\big(y_i\, f(\mathbf{x}_i)\big),
\end{equation}
where $\Phi(\cdot)$ is the cumulative distribution function of the standard normal distribution. The kernel hyperparameters $\theta$ and variational parameters $(\boldsymbol{\mu}_u, \mathbf{S}_u)$ are jointly learned by maximizing the evidence lower bound (ELBO):
\begin{equation}
\mathcal{L}_{\mathrm{ELBO}} = \sum_{i=1}^N \mathbb{E}_{q(f(\mathbf{x}_i))}\!\left[\log \Phi\!\big(y_i\, f(\mathbf{x}_i)\big)\right] - \mathrm{KL}\!\left[q(\mathbf{u}) \,\|\, p(\mathbf{u})\right].
\end{equation}
At prediction time, the class probability is obtained by integrating the probit link function against the predictive distribution $q(f(\mathbf{x}_\star))$, which yields:
\begin{equation}
p\big(y_\star=1 \mid \mathbf{x}_\star\big) = \int \Phi(f)\,\mathcal{N}\!\big(f \mid \mu_\star, v_\star\big)\, df = \Phi\!\left(\frac{\mu_\star}{\sqrt{1+v_\star}}\right).
\end{equation}

\section{Neural networks}

While Gaussian process models are highly interpretable, their expressiveness is constrained by the choice of kernel. To explore the limits of predictive performance on our dataset, we also train a deep neural network (NN) baseline.

The outputs of the convolutional module are flattened into a single feature vector, combining all histogram properties. Each histogram is represented by a vector of dimension $N_{\rm F}=64$, leading to an overall feature vector of size $N_{\rm F}\times N_{\rm histograms}$. Fully connected layers are then used to transform this high-dimensional representation into lower-dimensional feature embeddings, facilitating global interactions across all input properties. We use two fully connected layers, the first one with 150 neurons and the second with 32 neurons. For the classification task (SC vs. non-SC), the network concludes with a sigmoid activation function, producing a probability score suitable for binary classification. For the $T_c$ prediction task, the last output is not passed through a sigmoid, since it should return a continuous number. This process is illustrated in Fig.~\figref{fig:ML-workflow}{b}.

\section{Performance comparison}
\begin{figure}
    \centering
    \includegraphics[width=\linewidth]{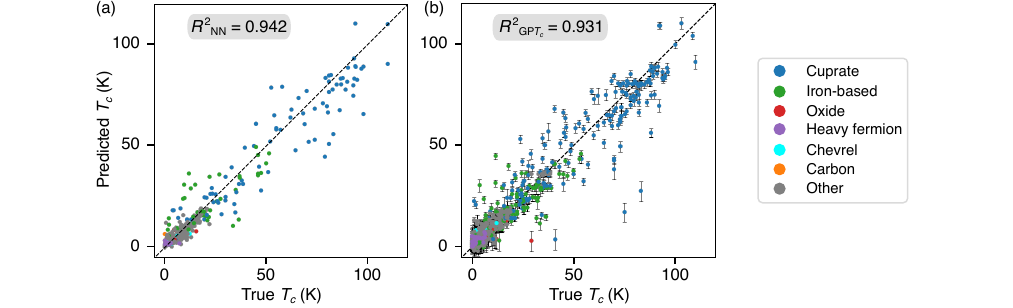}
    \caption{Performance comparison in $T_c$ prediction of (a)~the neural network and (b)~the Gaussian process, GP$T_c$. The models are almost on par.}
    \label{fig:performance_comparison}
\end{figure}

Figure~\ref{fig:performance_comparison} shows that the performance of our GP model, GP$T_c$, is very close to that of the neural network. While the NN yields $R^2=0.942$, GP$T_c$ gets to $R^2=0.931$ which is very close, while being interpretable.

\begin{figure}
    \centering
    \includegraphics[width=\linewidth]{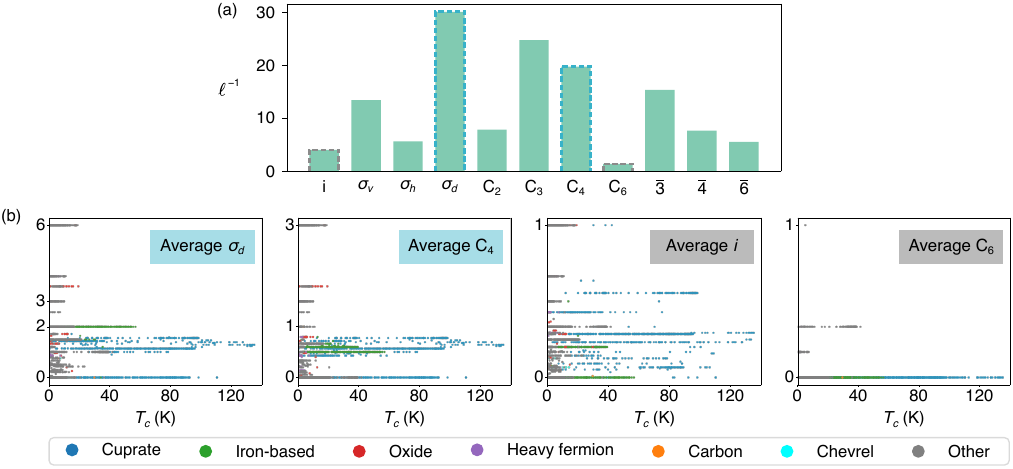}
    \caption{
    (a)~Inverse length scales of the 11 symmetry features, in $T_c$ prediction.
    (b)~$T_c$ vs. four exemplary average symmetry features. In the left plots, $\sigma_{d}$ and $C_{4}$ [shown as dashed blue lines in (a)], which are learned as highly predictive by the GP, show distinct shapes that partly differentiate between values of $T_c$. In the right plots, $i$ and $C_6$ [shown as dashed gray lines in (a)], which are learned as not predictive by the GP, exhibit either a large spread in $T_c$ ($i$) or almost no change in $T_c$ ($C_6$).
    }
    \label{fig:gp-interpretation}
\end{figure}

An unexpected outcome was the little significance the symmetry features carried in determining $T_c$ (see the feature ranking in Sec.~V.A). To gain insight into this, we trained a model with four of the best histogram features and all 11 symmetry features. We show the resulting GP length scales ($\ell_n$ in Eq.~(1) of the main text) associated with the 11 symmetry features in Fig.~\figref{fig:gp-interpretation}{a}. Notice that the absolute scale of the length scale $\ell$ for the graphlet features and the histogram features are different as they enter two separate kernels. Inspecting the actual distribution of the symmetry features, more predictive symmetry features ($\sigma_d$ and $C_4$) exhibit distinct distributions that correlate with $T_c$ trends. On the other hand, the two least predictive features either show a broad, uninformative spread with $T_c$ ($i$) or display almost no variation with $T_c$ at all ($C_6$); see Fig.~\figref{fig:gp-interpretation}{b}. All in all, given that the $T_c$ prediction task was based on superconducting materials, the fact that most of the materials share similar symmetry removed predictive power from symmetry features when it came to predicting $T_c$ among superconductors. The symmetry features, however, are more predictive of whether a material will superconduct.

\section{Identifying significant features}

\subsection{$T_c$ prediction}

Our GP regression results [Fig.~3(d) of the main text] show that the significant boost in performance comes from including second-order histogram features, i.e., going beyond simple averages. Adding third-order histogram features provides only a modest improvement on top of that. In light of this, and in order to reduce the combinatorial space, we limit our feature removal experiments to second-order histogram features + symmetry features.

Our feature removal experiment proceeds as follows. We start from the full set of $N=32$ features (21 second-order histogram features + 11 symmetry features), and train a $T_c$ prediction GP model. We then train $N$ different GP models, where each time a different one of the $N$ features is removed, so only the remaining $N-1$ features participate in the regression. For each model, we evaluate the performance using the $R^2$ score on the test set. Identifying the model with the highest test $R^2$ score singles out the least predictive feature: the one that has been removed in that particular model. This leaves us with a set of $N-1$ features, that are the most predictive subset of the original set of $N$ features.

We then proceed iteratively: starting from $N-1$ features that were most predictive in the previous iteration, we train $N-1$ models (each with a different one of the $N-1$ features is removed), and find the model with the highest test $R^2$ score. This allows us to remove an additional feature, leaving us with $N-2$ features. This process repeats iteratively all the way to keeping just one feature. We track the performance (test $R^2$ score) of the best models along this feature removal process in Fig.~3(e) of the main text. We find that the best overall model yields $R^{2}_{\rm opt}=0.933$, and keeping just four features yields $R^{2}_{4}=0.922$, which is quite close to $R^{2}_{\rm opt}$ (for reference, the most predictive single feature gives $R^{2}_{1}=0.864$). It is also apparent from Fig.~3(e) of the main text that the curve starts to flatten around four features, with the additional improvements from including extra features getting smaller and smaller.

We list here the features by the predictiveness according to the feature removal analysis, sorted from the most predictive one to the least predictive one:
\begin{enumerate}
\item Electron affinity difference
\item Inter-atomic distance
\item Total number of valence electrons mean
\item Column in the periodic table mean
\item Number of d valence electrons mean
\item Number of p valence electrons mean
\item Atomic weight mean
\item Number of s valence electrons mean
\item Vertical mirror plane
\item 6-fold rotation axis
\item Pauling electronegativity mean
\item 6-fold rotoinversion axis
\item Covalent radius difference
\item 4-fold rotoinversion axis
\item Number of d valence electrons difference
\item Total number of valence electrons difference
\item 4-fold rotation axis
\item 3-fold rotation axis
\item Ionization potential mean
\item Dihedral mirror plane
\item 3-fold rotoinversion axis
\item Number of p valence electrons difference
\item Number of s valence electrons difference
\item Pauling electronegativity difference
\item Covalent radius mean
\item Column in the periodic table difference
\item Horizontal mirror plane
\item Atomic weight difference
\item Ionization potential difference
\item Inversion center
\item 2-fold rotation axis
\item Electron affinity mean
\end{enumerate}

This finding motivated us to examine all possible combinations of four features: the combinatorial space is $\binom{32}{4}=35,960$, which is large but reasonable. See the database for the outcome of the combinatorial search.

\subsection{Classification}
Following the same feature pruning procedure we used for $T_c$ prediction, we train GP models to identify the most predictive features for SC / non-SC classification. The features are listed here in order from the most predictive one to the least predictive one for classification:
\begin{enumerate}
\item Number of d valence electrons difference
\item Ionization potential mean
\item Atomic weight mean
\item Dihedral mirror plane
\item 3-fold rotation axis
\item Total number of valence electrons difference
\item Number of s valence electrons mean
\item Electron affinity difference
\item Number of p valence electrons difference
\item Number of p valence electrons mean
\item Atomic weight difference
\item Inversion center
\item Pauling electronegativity mean
\item Covalent radius difference
\item Number of s valence electrons difference
\item Total number of valence electrons mean
\item Column in the periodic table difference
\item Column in the periodic table mean
\item Inter-atomic distance
\item 4-fold rotation axis
\item Vertical mirror plane
\item 3-fold rotoinversion axis
\item 6-fold rotation axis
\item Ionization potential difference
\item Number of d valence electrons mean
\item Horizontal mirror plane
\item 4-fold rotoinversion axis
\item Covalent radius mean
\item 2-fold rotation axis
\item Pauling electronegativity difference
\item 6-fold rotoinversion axis
\item Electron affinity mean
\end{enumerate}

\begin{figure}
    \centering
    \includegraphics[width=0.5\linewidth]{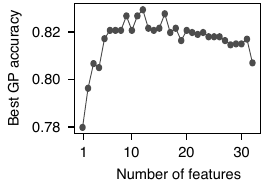}
    \caption{Performance of the GP classifier when only subsets of features are included. The model's performance begins to degrade when the number of features falls below 10.}
    \label{fig:classifier_pruning}
\end{figure}

The feature-pruning experiment on the classifier shows that more histogram features are necessary for classifier predictions. Specifically,
Fig.~\ref{fig:classifier_pruning} shows that achieving full performance requires approximately 10 histogram features (see Sec.~V.B).
Atomic weight mean and EA difference stand out as predictive for both classification and $T_c$ regression, and apart from them the predictive features for the two tasks have little overlap, with the symmetry features being much more predictive for classification than for regression (see Sec.~V).
Our observation that features predicting whether a material superconducts are distinct from the features predicting $T_c$ is in line with the observations in an early manual statistical analysis using elemental features~\cite{hirsch_correlations_1997}.

\subsection{Physical basis and mechanism-agnostic relevance of electron affinity difference}

Our feature-removal analysis identifies the second-order electron-affinity (EA) difference as the single most predictive graphlet feature for $T_c$ regression, and it remains among the most informative features for superconductivity classification. This finding invites a physical interpretation: why should the distribution of EA differences between neighboring atoms be so closely tied to superconductivity?

The electron affinity of an atom measures its tendency to accept an additional electron and is therefore a direct indicator of its bonding character. The EA difference between neighboring sites provides an inexpensive, purely chemical window into local electronic structure---small differences signal covalent bonding, large differences ionic---without requiring band structure calculations. 
Since the electronic energy scales and conduction electron's coupling with the lattice are all governed by the local bonding character, it is natural that EA-difference distributions emerge as a powerful predictor of $T_c$ across superconductor families. For transition-metal oxides, the connection between the key electronic energy scale and EA difference can be made precise: the EA difference maps directly onto the charge-transfer energy $\Delta$ in the Zaanen--Sawatzky--Allen framework~\cite{ZaanenSawatzkyAllen1985}, where $\Delta$ is set by the relative cation and anion orbital energies in the crystal environment. While this notion is  specific to compounds with an electronegative anion, the broader statement---that EA differences track local bonding character, and that bonding character sets the relevant electronic energy scales---is not restricted to oxides, and it is that broader statement on which our descriptor rests. The EA-difference and interatomic-distance distributions in our graphlet features therefore encode the local chemical and structural factors that govern $\Delta$ and related electronic energy scales~\cite{WatkinsCheethamSeshadri2025}.

The physical significance of this connection becomes particularly transparent in the cuprates, where superconductivity is widely associated with spin fluctuations whose characteristic energy scale is the antiferromagnetic superexchange coupling $J$~\cite{Keimer_Cuprates_2015}. The key insight is that $J$ is not independent of local bonding chemistry---it is controlled by it. In the three-band Emery model~\cite{Emery1987}, the Cu--O planes are described by the Hamiltonian
\begin{equation}
H=\epsilon_d\sum_{i\sigma}n^d_{i\sigma}+\epsilon_p\sum_{j\sigma}n^p_{j\sigma}
+t_{pd}\sum_{\langle ij\rangle\sigma}\eta_{ij}\left(d^\dagger_{i\sigma}p_{j\sigma}+\mathrm{h.c.}\right)
+U_d\sum_i n^d_{i\uparrow}n^d_{i\downarrow},
\end{equation}
where $d_{i\sigma}$ ($p_{j\sigma}$) annihilates an electron with spin $\sigma$ on Cu site $i$ (O site $j$), $\epsilon_d$ and $\epsilon_p$ are on-site orbital energies, $t_{pd}$ is the Cu--O hopping, $\eta_{ij}$ is the orbital-phase sign, and $U_d$ is the Cu on-site Coulomb repulsion. With the charge-transfer energy $\Delta=\epsilon_d-\epsilon_p$, virtual Cu--O processes generate superexchange~\cite{ZhangRice1988,ZaanenOles1988}:
\begin{equation}
J=\frac{4t_{pd}^{4}}{\Delta^{2}}
\left(\frac{1}{U_d}+\frac{1}{\Delta}\right).
\end{equation}
Stronger Cu--O covalency (smaller $\Delta$, larger $t_{pd}$) enhances $J$, and the EA difference between a transition-metal site and its ligand is a direct proxy for $\Delta$. Our descriptor captures this information as a \emph{distribution} over all nearest-neighbor pairs, encoding the bonding heterogeneity of the crystal---for instance, distinguishing the covalent Cu--O bonds within CuO$_2$ planes from the ionic bonds in the charge-reservoir layers. More generally, superconducting families need not share a cuprate-like mechanism: local chemistry and bonding geometry set the electronic energy scales relevant to each family's pairing interaction, whether phononic or magnetic. The EA-difference distribution therefore provides a compact, transferable predictor by encoding bonding character upstream of the specific mechanism.

\section{Filtering of new predictions}

\subsection{Feature Construction and Prediction Pipeline for ICSD}

To extend the model beyond the 3DSC dataset and evaluate superconductivity across the broader landscape of known inorganic materials, we generated graphlet and symmetry features for most entries in the ICSD. ICSD provides crystallographic data for more than 225{,}000 experimentally synthesized inorganic compounds, offering a comprehensive platform for large-scale screening. Using the featurization workflow described in the main text (and detailed in SM Sec.~I--II), we processed all ICSD entries with machine-readable CIFs to generate graphlet features for approximately 185{,}000 materials.

Graphlet features were computed following the same protocol as we did with the materials within the 3DSC. First-, second-, and third-order graphlets were constructed to capture local chemical environments and structural motifs through distributions of elemental descriptors, interatomic distances, and bond angles (see Fig.~2(a) and SM Sec.~I.C). For ICSD screening, we primarily relied on second-order graphlets together with symmetry features, since these were shown to retain the vast majority of predictive power while limiting model complexity. To ensure compatibility with the trained GP models, all ICSD histogram features were standardized using the same bin centers defined during 3DSC preprocessing. This alignment preserves a consistent feature space for EMD-based kernel comparisons.

Predictions from the GP classifier and $\mathcal{GP}$-$T_c$ for each ICSD ID formed the starting point of our analysis.

\subsection{Post-processing and Filtering of ICSD Predictions}

The full set of approximately 185{,}000 predictions was subsequently filtered to identify promising superconducting candidates and prioritize materials for follow-up analysis.

\subsubsection{$T_c > 10$~K}
In this regime, the $\mathcal{GP}$-$T_c$ exhibits relatively low uncertainty [Fig.~3(b--c) of the main text], making its output a reliable basis for screening. We focus on materials with
\begin{enumerate}
    \item the predicted $T_c > 10$~K,
    \item classification probability $> 0.5$.
\end{enumerate}
To focus on unexplored chemical space, we removed materials belonging to the well-established high-$T_c$ families: cuprates and iron-based superconductors. The corresponding filtered candidate set is shown in Fig.~4e of the main text.

\subsubsection{$2\,\text{K} < T_c < 10$ K}
There are many more predictions for $T_C<10$ K though likely they are of less interest to the community. To establish the feasibility of GP$T_c$ driven synthesis and confirmation of new superconducting material, we narrowed the candidate pool using pragamatic criteria. To avoid the use of a dilution refrigerator and aim for a most likely success within the material space that Schoop lab can readily access, we required the following criteria:
\begin{enumerate}
    \item We considered $2\,\text{K} < \text{Predicted } T_c < 10\,\text{K}$.
    \item We required the classification probability $> 0.75$.
    \item We removed compounds containing O, N, F (outside the expertise of the lab), Os, Be ,Tl, As (because of toxicity), Mn, Gd (usually lead to magnetism), H (hard to characterize), and radioactive elements.
    \item We only kept stoichiometric (undoped) compounds with up to three distinct elements, to make growing more straightforward.
    \item We removed compounds who chemical formula contains 10 or more of the same elements, since it would be hard to establish that the correct composition was indeed grown.
    \item We removed compounds that have a large band gap ($>0.1\,{\rm eV}$), either according to density functional theory (as recorded in the Materials Project~\cite{jain_commentary_2013-1,munro_improved_2020}) or according to our automated literature search (see the next section). While such insulators could be interesting as potential SCs upon doping, our initial proof of concept quest focused on undoped SCs.
    \item We removed compounds containing a known elemental SC whose predicted $T_c$ is within 1~K of the elemental SC's $T_c$, since these could be mistaken for elemental SCs with impurities.
\end{enumerate}

The resulting candidate list was forwarded to the Edison automated literature-precedence search agent to identify related compounds, known superconductors not represented in 3DSC, or materials with no documented history of superconductivity (see the next section for details).
After applying these filters, we obtained a list of 173 potentially new, practically synthesizable SCs.

\subsection{Literature search for known superconductivity}

Since the SuperCon database is not exhaustive and dated, some of the SCs predicted by GP$T_c$ had been already confirmed. We therefore conducted a comprehensive literature search to identify any precedent.
For materials in the filtered prediction list, we queried for known $T_c$ with the prompt: \texttt{Is [FORMULA] a known superconductor? If so, what is the highest reported Tc and what is the paper source?}. Here, \texttt{[FORMULA]} is replaced with the chemical formula of the material.

We used Edison Scientific's Precedent search agent for literature search (\url{https://edisonscientific.gitbook.io/edison-cookbook/edison-client}). With this prompt, Precedent (1)~retrieved publications from the literature that are relevant to the material, (2)~perused articles using a large language model, and (3)~synthesized a report detailing prior superconductivity research on the material. The agent ensures an exhaustive literature search, i.e., maximizes knowledge recall, by leveraging multiple search rounds. Each time, the agent gathers more focused search results and incorporates information from previous rounds. Based on the PaperQA2 framework of~\citet{skarlinski2024language}, the agent is optimized to answer questions such as ``Has anyone done X before?'' The full pipeline is proprietary.

For all materials, the Precedent search agent created reports of search results, including the reported $T_c$ and publication references, if any. Even when no experimental $T_c$ was reported, the reports often cited closely related superconducting materials, electronic or magnetic phase information, and insulating properties. The summaries of the properties and sources we list in our database will be valuable to the community in evaluating our predictions.

\subsection{Synthesis and characterization of PtPb$_3$Bi}

To validate our model's predictive power, we synthesized PtPb$_3$Bi, which $\mathcal{GP}$-$T_c$ predicts to have $T_c$ of 2.93~K, with a confidence of 0.756.
This was a pragmatic choice as both Pb and Bi form deep eutectics with Pt, offering a low-temperature route to grow large single-crystals for experimental verification of its superconducting properties.
Remarkably, magnetometry shown in Fig.~5(c--d) of the main text establishes that PtPb$_3$Bi indeed undergoes a type-II superconducting transition with $T_c\approx2.98$~K. The temperature dependence of our magnetic susceptibility shows that the demagnetized-corrected, field-cooled curve reaches a value of 4$\pi\chi_v\sim-1.06$, suggesting a newly discovered, phase-pure superconductor.

\subsubsection{Materials and Synthesis}

Samples of PtPb$_3$Bi were synthesized using a self-flux method utilizing a Canfield crucible. In the bottom crucible, 7 at\% Pt (powder, Sigma Aldrich, 99.995\%), 69.75 at\% Pb (shot, Alfa Aesar, 99.999$\%$), and 23.25 at\% Bi(pieces, Sigma Aldrich, 99.999\%) were weighed out and combined, totaling 2g. The Canfield crucible setup was then place in a quartz tube, flushed with Ar 3 times, and sealed under vacuum at 80 mTorr. The sealed quartz tube was then placed in an induction furnace, ramped to 500$^{\circ}$C over 5 hours and held at 500$^{\circ}$C for 48 hours to ensure homogenization of precursors. The furnace was subsequently slowly cooled to 300$^{\circ}$C at a rate of 4K/hour, and held at 300$^{\circ}$C for 48 hours before centrifuging. The sample resulted in an ingot consisting of large metallic crystals, within a Pb matrix. Crystals used in measurements were then mechanically separated from the Pb matrix.

\subsubsection{Single Crystal X-ray diffraction}

Single crystal X-ray diffraction measurements were performed at room temperature using a Rigaku XtaLAB Synergy-S/i diffractometer equipped with a Mo K$_{\alpha1}$ ($\lambda$ = 0.71073~\AA), micro-focus sealed-tube X-ray source and a graphite monochromator. Integration was carried out using CrysAlis$^{\text{Pro}}$ software, with numerical absorption correction based on gaussian integration over a multifaceted crystal model. Full structural refinements on F$^2$ were performed in JANA2020.

Unit cell finding finds the previously reported cell as a tetragonal cell with $a$ = 11.4557(8)~\AA\ and $c$ = 4.0836(3)~\AA. Refinement proceeds smoothly in the previously reported space group of tP4$_2$/$m n m$.

Crystallographic refinement information, atomic coordinates, anisotropic parameters, and distances are given in Tables~\ref{CI_PPB}-\ref{D_PPB}.

\begin{center}
\begin{table*} [hbt!]
\caption{Crystallographic Information for {PtPb$_3$Bi}.}
 \centering
 \begin{tabular}{l c}
\hline
Refined Composition & PtPb$_3$Bi \\
\hline
Crystal Dimension (mm) & 0.033 $\times$ 0.036 $\times$ 0.134  \\
Radiation Source, $\lambda$ (\AA) & { Mo K$_{\alpha}$, 0.71073 }   \\
Absorption Correction &  Gaussian   \\
Data Collection Temperature (K)   &  293.15(10) \\
Space Group &  P 4$_2$/$m n m$   \\
$a$ (\AA) &  11.4557(8)   \\
$c$ (\AA) &  4.0836(3) \\
Cell Volume (\AA$^{3}$)  &  535.90(6) \\
Absorption Coefficient (mm$^{-1}$) & 152.65 \\
$\theta_{\rm min}$ , $\theta_{\rm max}$ (deg) &  2.51, 39.97 \\
Refinement Method & F$^{2}$  \\
R$_{int}$(I\textgreater{}3$\sigma$, all)   &  30.22, 31.85    \\
Number of Parameters     &  19    \\
Unique Reflections (I\textgreater 3$\sigma$, all)  & 1007, 543  \\
R(I\textgreater{}$3\sigma$), R$_{w}$(I\textgreater{}$3\sigma$)   &   5.19, 9.46    \\
R(all), R$_{w}$(all)      & 11.55, 10.16   \\
S(I\textgreater{}$3\sigma$), S(all)   & 2.0887, 1.6347   \\
$\Delta\rho_{\rm max}$ , $\Delta\rho_{\rm min}$ (e \AA$^{-3}$) & 8.77, -7.87  \\
\hline
\label{CI_PPB}
\end{tabular}
\end{table*}
\end{center}

\begin{center}
\begin{table*}[hbt!]
\caption{Refined atomic coordinates for PtPb$_3$Bi.}
\begin{tabular}{@{}lcccccc@{}}
\hline
Site & Wyckoff Position & x   & y    & z          & Occupancy &  \\
\hline
Pt1 & 4f & 0.58675(6) & 0.58675(6) & 0    & 1 \\
Pb1 & 4f & 0.39768(6) & 0.60232(6) & 1/2  & 1 \\
Pb2 & 8i & 0.50243(7) & 0.82996(7) & 0    & 1 \\
Bi1 & 4g & 0.70728(6) & 0.70728(6) & -1/2 & 1 \\
\hline
\label{AC_PPB}
\end{tabular}
\end{table*}
\end{center}

\begin{center}
\begin{table*}[hbt!]
\caption{Refined anisotropic displacement parameters for PtPb$_3$Bi.}
\begin{tabular}{lcccccc}
\hline
Site & U$_{11}$ & U$_{22}$   & U$_{33}$    & U$_{12}$ & U$_{13}$         & U$_{23}$   \\
\hline
 Pt1 & 0.0226(3) & 0.0226(3) & 0.0222(5) & -0.0030(3)  & 0 & 0  \\
 Pb1 & 0.0217(3) & 0.0217(3) & 0.0213(5) & 0.0010(3)   & 0 & 0  \\
 Pb2 & 0.0263(3) & 0.0235(3) & 0.0332(4) & 0.0015(2)   & 0 & 0  \\
 Bi1 & 0.0209(3) & 0.0209(3) & 0.0218(5) & -0.0005(3)  & 0 & 0  \\
\hline
\label{AP_PPB}
\end{tabular}
\end{table*}
\end{center}

\begin{center}
\begin{table*}[hbt!]
\caption{Selected interatomic distances for PtPb$_3$Bi.}
\begin{tabular}{lccc}
\hline
 Site & Neighbor & Multiplicity   & Distance (\AA)  \\
\hline
  Pt1 & Pt1 & 1 & 2.8109(10) \\
  Pt1 & Pb1 & 4 & 2.9819(8)  \\
  Pt1 & Pb2 & 1 & 2.9488(11) \\
  Pt1 & Bi1 & 2 & 2.8252(7)  \\
  Pb1 & Pb1 & 1 & 3.3153(10) \\
  Pb1 & Pb2 & 4 & 3.5227(9)  \\
  Pb1 & Bi1 & 2 & 3.7449(11) \\
  Pb1 & Bi1 & 2 & 3.6992(8)  \\
  Pb2 & Pb2 & 4 & 3.4292(9)  \\
  Bi1 & Pb2 & 2 & 3.4080(11) \\
  Bi1 & Pb2 & 2 & 3.4133(9)  \\
\hline
\label{D_PPB}
\end{tabular}
\end{table*}
\end{center}

\subsubsection{Energy Dispersive X-ray Spectroscopy}
\begin{figure*}[ht!]
    \resizebox{0.7\textwidth}{!}{\includegraphics{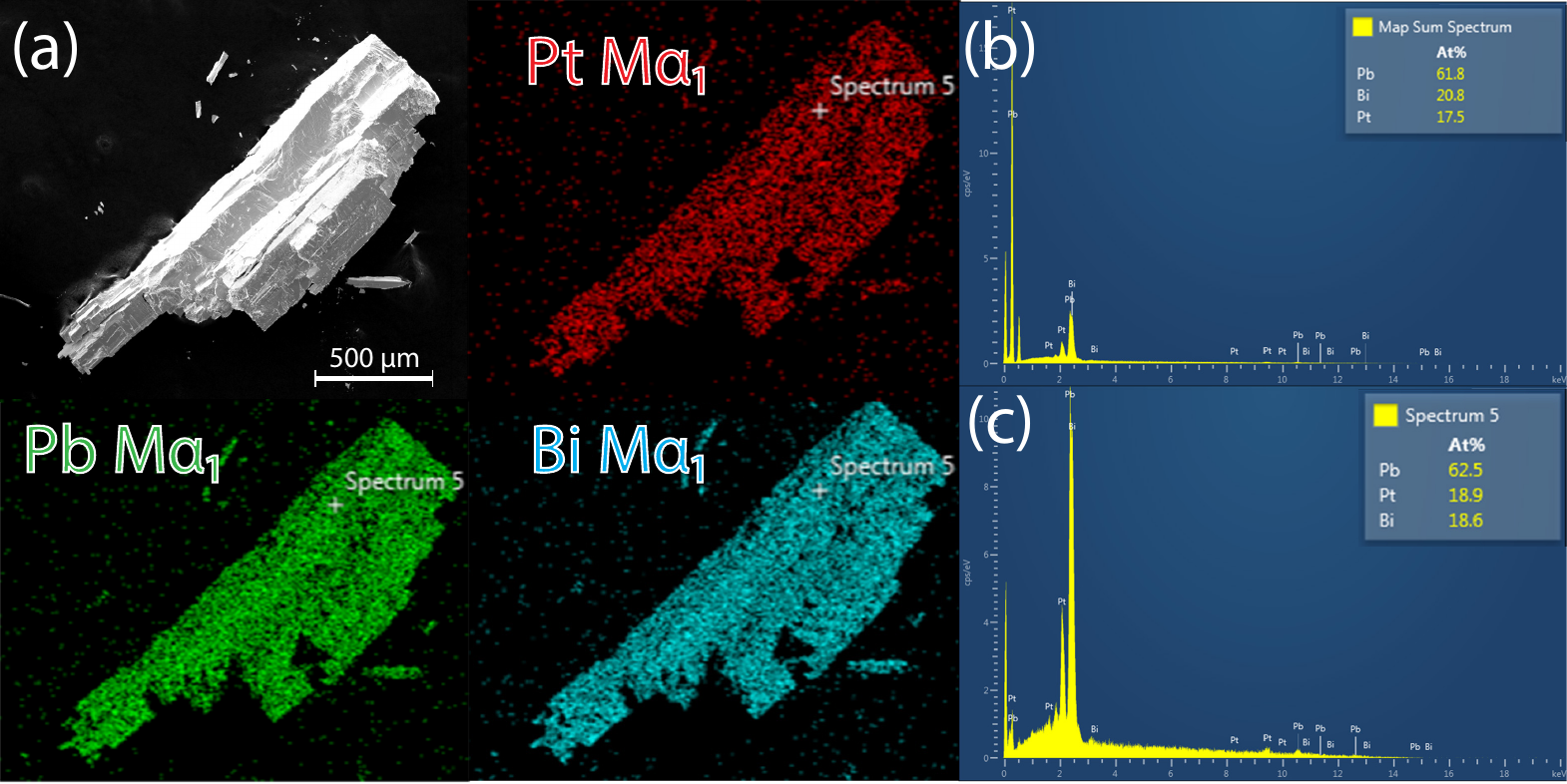}}
\caption[Chemical composition analysis of PtPb$_3$Bi via EDX]{\textbf{Chemical composition analysis of PtPb$_3$Bi via EDX} \textbf{(a)} Elemental mapping of constituent elements in PtPb$_3$Bi, showing homogeneous signal throughout the mapped crystal. \textbf{(b)} Map-sum and \textbf{(c)} Point spectra of the picked crystal, confirming the stoichiometry of the sample as PtPb$_3$Bi.}
\label{EDX}
\end{figure*}

Crystals of PtPb$_3$Bi that were mechanically removed from an ingot were characterized using energy-dispersive x-ray spectroscopy in a Quanta enviornmental scanning electron microscope equipped with an Oxford EDX detector. Compositional analysis and sample homogeneity is shown in Fig.~\ref{EDX}.

\subsubsection{Superconductivity Characterization}
\begin{figure*}[ht!]
    \resizebox{0.6\textwidth}{!}{\includegraphics{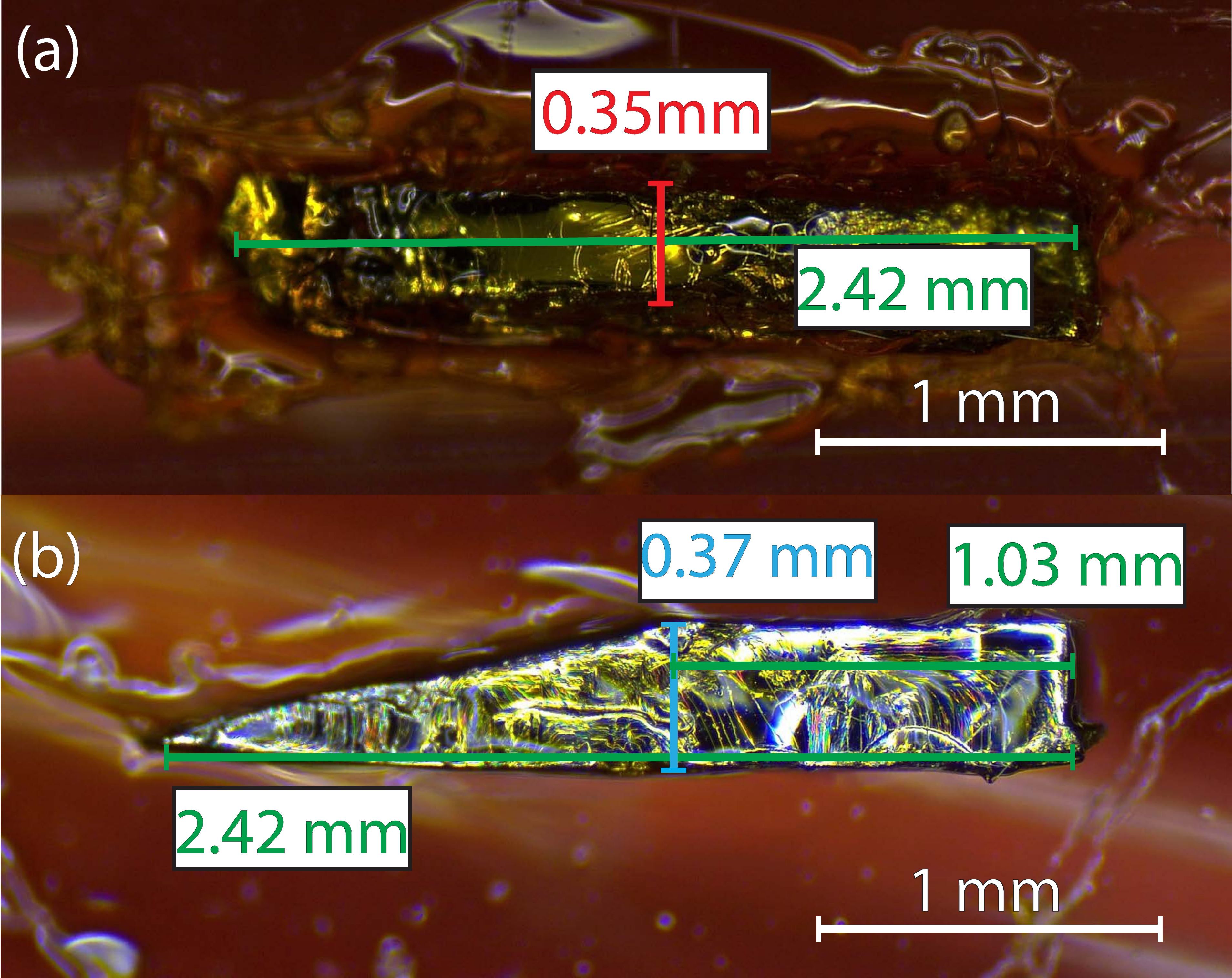}}
\caption[Geometry of crystal measured in magnetic susceptibility measurements]{\textbf{Geometry of crystal measured in magnetic susceptibility measurements} \textbf{(a)} Crystal measurements of A (red) and C (green). \textbf{(b)} Crystal measurements of B (blue) and C (green).}
\label{Demag}
\end{figure*}

Data for DC magnetic susceptibility measurements utilized a Quantum Design Magnetic Properties Measurement system. A single crystal, shown at the inset of Fig.~\ref{Demag}, was mounted on a quartz holder with GE varnish. The sample was mounted with the c-axis aligned parallel to the applied magnetic field to minimize demagnification effects. Data points for M-T sweeps were taken at $\sim0.005$~K intervals, with field-cooled data taken with an applied field of $\sim5$~Oe.

The sample volume was determined by the total mass of the sample, 3.04~mg, and the nominal density of the sample, 12.712~g/cm$^3$. The volume-weighted measured total moment was then deduced by dividing the total magnetic moment by this volume.

We then corrected the internal field due to demagnetization factors
\begin{equation}
    H_{\rm int} = H_{\rm app} - 4\pi NM,
\end{equation}
where $H_{\rm int}$ is the internal magnetic field, $H_{\rm app}$ is the apparent magnetic field, N is the demagnetization factor that is geometrically dependent, and M is the magnetization of the sample. We estimate our crystal as a rectangular prism geometry with the c-axis parallel to the applied magnetic field~\cite{Lygouras_2025, Prozorov_2018}, in this case

\begin{equation}
    N \approx \frac{4AB}{4AB+3C(A+B)},
\end{equation}
where $A$ and $B$ are lengths of the crystal perpendicular to the applied magnetic field, and $C$ is the length of the crystal parallel to the crystal field. For the demagnetization factor we note that our measured crystal, shown in Fig.~\ref{Demag} is not a perfect rectangular prism. For our calculation of N, we estimate the values of $A=0.35$~mm, $B=0.37$~mm and $C=1.74$~mm, where the length of $C$ is estimated for a rectangular prism that conserves the total volume of the sample. This results in a demagnification factor of $N\approx0.1211$. This estimation of C should slightly overestimate the true value of N. The slight overshooting of the expected phase pure superconducting value (-1) can be attributed to the extra diamagnetic signal of the GE varnish and quartz holder.

\begin{figure*}[ht!]
    \resizebox{0.5\textwidth}{!}{\includegraphics{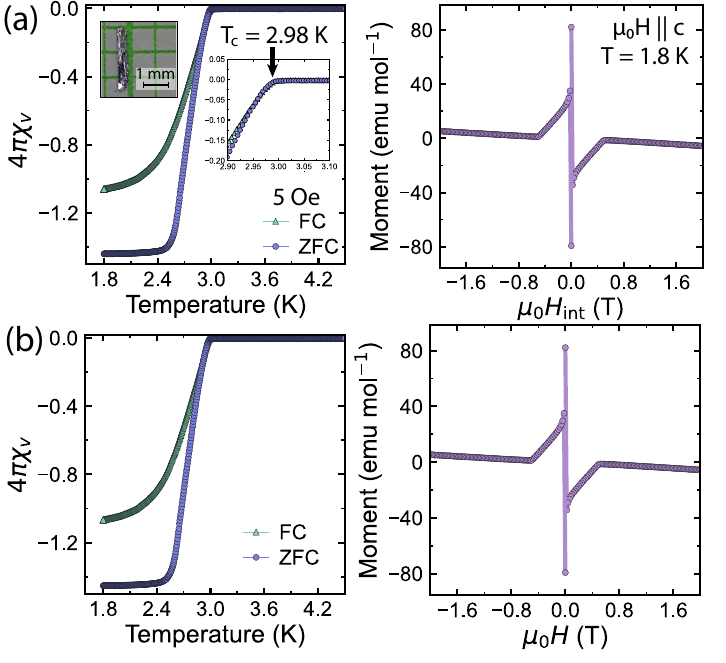}}
\caption[Demagnetization correction for M-T and M-H sweeps]{\textbf{Demagnetization correction for M-T and M-H sweeps} \textbf{(a)} M-T and M-H sweeps of PtPb$_3$Bi with applied demagnetization correction N = 0.1211. \textbf{(b)} M-T and M-H sweeps of PtPb$_3$Bi without any demagnetization correction.}
\label{DemagComp}
\end{figure*}

In Fig.~\ref{DemagComp}, we show the magnetic susceptibility of both the uncorrected and corrected M-T and M-H runs.

Alternating current electronic transport measurements were performed using a Quantum Design Physical Property Measurement System. Gold wires were connected in a four probe geometry using conducting silver paste (Dupont 4929N) to the same crystal measured in our magnetization data.

\begin{figure*}[ht!]
    \resizebox{0.7\textwidth}{!}{\includegraphics{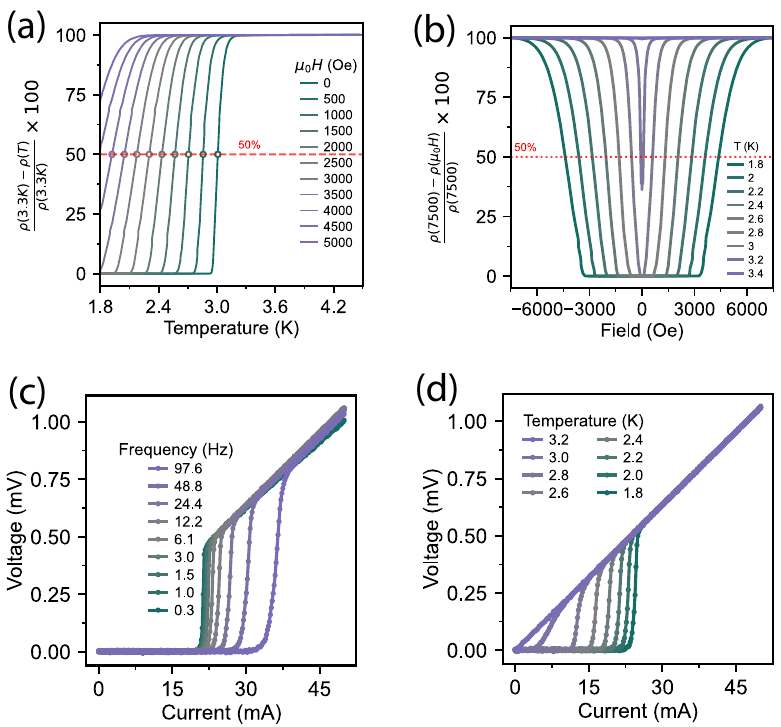}}
\caption[Electronic transport data for PtPb$_3$Bi.]{\textbf{Electronic transport data for PtPb$_3$Bi.} \textbf{(a)} Temperature dependent resistivity at different applied field confirm a bulk superconductivity drop at 3.0~K, consistent with our magnetization data. \textbf{(b)} Field dependent resistivity (magnetoresistivity) at different temperatures. \textbf{(d)} Temperature- and \textbf{(e)} frequency-dependent $I-V$ curves consistent with a type-II superconducting transition.}
\label{ETO_Data}
\end{figure*}

Field dependent $\rho$-T curves were taken under different applied fields at intervals of 500 Oe (Fig.~\ref{ETO_Data}(a)). All fields show a drop to 0 resistivity at low temperatures. At zero applied field, the resistivity drops by 50\% at 3.009~K, further confirming the bulk superconducting transition shown in our $M-T$ curves. Further, the transition width, which we define as the change in temperature between the 90\% and 10\% drop in resistivity, is only 0.111~K, indicating that our sample is of high crystalline quality. Upon increasing the applied field, the transition temperature for PtPb$_3$Bi decreases and the transition width increases, as expected for a type-II superconductor.

The isotherms of $\rho$-$\mu_0H$ (Fig.~\ref{ETO_Data}(b)) also display expected behavior for superconductivity, where increasing temperature reduces the upper critical field towards the normal state. Utilizing the data from both $\rho$-T and $\rho$-$\mu_0H$ experiments, we then plot a phase diagram for our upper critical field (Fig.~\ref{ETO_Data}(c)). A simple Werthamer-Helfand-Hohenberg (WHH) fit to this data suggests an estimate to the upper critical field as $\mu_0H_{c2}(0)=$0.6672~T. This upper critical field falls far below the Pauli limit, $H_P\approx1.86\times3.0\approx5.6~T$, suggesting an orbital-pair breaking mechanism.
We also note that this value of the critical field is an order of magnitude higher than that of elemental Pb, for which $\mu_0H_{c2}(0)=$0.0803~T~\cite{ChaninTorre1972}, suggesting that the superconductivity in this compound is not a trivial consequence of its Pb content.

Temperature- and frequency-dependent $I-V$ curves are shown in Fig.~\ref{ETO_Data}(d) and (e), both displaying expected behavior for superconductivity. At 1.8~K an AC current frequency of 97.6~Hz has the highest measured critical current of $\sim$33~mA. Lower AC frequencies result in lower critical currents, consistent with the relaxation of vorticies only seen in type-II superconductors.

\begin{figure*}[ht!]
    \resizebox{0.8\textwidth}{!}{\includegraphics{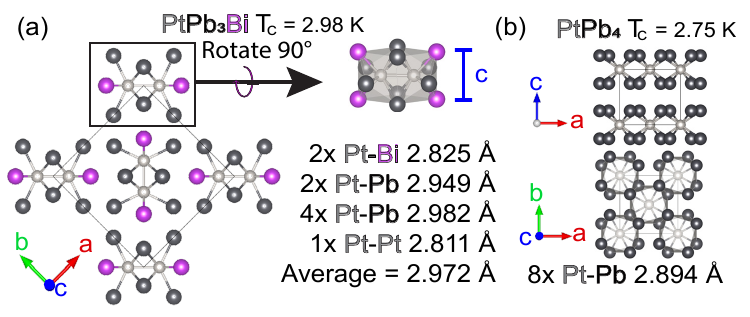}}
\caption[Structural comparison of PtPb$_3$Bi and PtPb$_4$.]{\textbf{Structural comparison of PtPb$_3$Bi and PtPb$_4$.} \textbf{(a)}~PtPb$_3$Bi, a quasi-one-dimensional stucture type with no other reported structural variants. The quasi-one-dimensional units are built up of dimerized bi-capped trigonal prismatic units, stacked along the c-axis. \textbf{(b)}~PtPb$_4$, built up of edge-sharing square antiprisms in a quasi-two-dimensional structure type. Recent characterization has classified it as a superconductor with $T_c\approx2.75$~K~\cite{Xu_2021, Wang_2021, GarciaTalavera_2025}.}
\label{Structural}
\end{figure*}

It is interesting to note that newly confirmed superconductor PtPb$_3$Bi  shares key 2nd-order graphlet features we have identified, electron affinity differences, and interatomic distances, with PtPb$_4$ ($T_c\sim2.75$~K) that has been recently characterized ~\cite{Xu_2021, Wang_2021, GarciaTalavera_2025}.
The two materials are markedly different in their structures, as shown in Fig.~\ref{Structural}. Specifically,  PtPb$_4$ hosts edge-sharing [PtPb$_8$] square antiprisms in quasi-two-dimensional layers, while PtPb$_3$Bi hosts face-sharing dimerized [Pt$_2$Pb$_8$Bi$_4$] bicapped trigonal prismatic units, stacked along the c-axis forming a quasi-one-dimensional structure type. Critically, PtPb$_3$Bi forms in its own unique structure type with no other reported variants to the best of our knowledge.
Hence, the prediction of superconductivity in PtPb$_3$Bi would have escaped any structure-focused approach, including a focus on structural motifs. On the other hand, the GP$T_c$'s success in correctly predicting $T_c$ for PtPb$_3$Bi to be close to that of PtPb$_4$ demonstrates the power of our four features in Fig 4(a).

\section{Family-specific analysis}

\subsection{Family-resolved performance}

To address whether our bonding-descriptor framework applies equally well to conventional and unconventional superconductors, we evaluate the model's performance resolved by material family. Specifically, we analyze the three largest subsets in our test data: ``Other'' (predominantly conventional phonon-mediated SCs), cuprates, and iron-based superconductors. Because standard metrics like $R^2$ depend strongly on the total $T_c$ variance of each subset, comparing raw $R^2$ values across families introduces significant bias. Instead, we fit a linear regression $T_c^{\text{pred}} = a T_c^{\text{true}} + b$ for each family, where ideal performance corresponds to unit slope ($a=1$) and zero intercept ($b=0$). To compare intercepts fairly across families with vastly different $T_c$ scales, we define a dimensionless intercept $b' = b / \text{IQR}$, normalizing $b$ by the interquartile range of $T_c$ within each family.

Evaluating the full test set yields an overall slope of $a=1.00$ and intercept $b=0.15~\text{K}$ ($b'=0.007$). When resolved by family, we obtain $a = 0.98, 0.95, 0.90$ and $b' = -0.02, 0.08, 0.11$ for the ``Other'', cuprate, and iron-based families, respectively. Crucially, the 95\% confidence intervals for the slope $a$ in all three families encompass the identity line ($a=1$). This demonstrates that there is no statistically significant difference in predictive accuracy across conventional, cuprate, and pnictide/chalcogenide families. The model learns a family-agnostic relationship between local bonding features and $T_c$, demonstrating that its predictive power is not restricted to conventional BCS systems.

\subsection{Doping dependence in the cuprates}

As a further test on strongly correlated unconventional systems, we evaluate $\mathcal{GP}$-$T_c$ on the doping-dependent phase diagram of the $RE\text{Ba}_2\text{Cu}_3\text{O}_{6+x}$ ($RE$-123) cuprate family ($RE = \text{Y, Eu, Gd, Tm, La}$). This family presents two stringent challenges for a structure- and element-based representation. First, $T_c$ is tuned across its full dome by subtle changes in oxygen stoichiometry $x$ at fixed cation composition, requiring the model to order nearly identical chemical formulas along the doping axis. Second, near optimal doping, $T_c$ remains localized around 90--95~K across a wide variation of $RE^{3+}$ ionic radii, meaning the model cannot rely simply on rare-earth elemental proxies.

\begin{figure}
    \centering
    \includegraphics[width=0.85\linewidth]{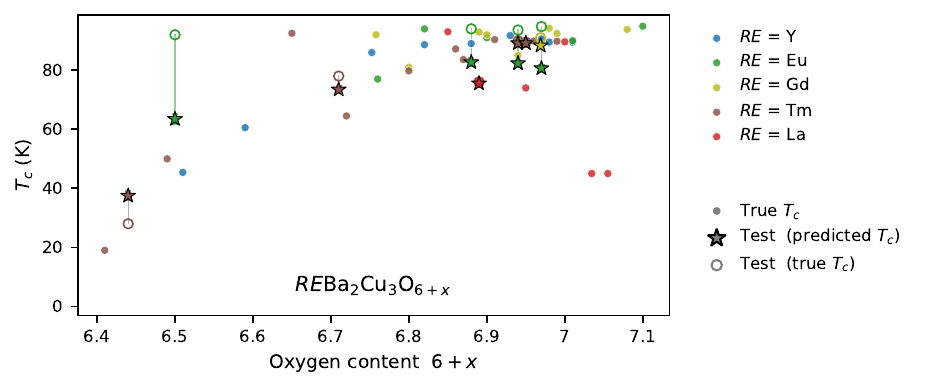}
    \caption{
    Critical temperature of the $RE$Ba$_2$Cu$_3$O$_{6+x}$ family as a function of oxygen content, for $RE=$~Y, Eu, Gd, Tm, and La (denoted by colors).
    Filled circles are experimental $T_c$ values for entries outside our test set.
    For the ten test-set entries, stars are the $T_c$ predicted by ${\cal GP}$-$T_c$, open circles the corresponding measured $T_c$, and the connecting segment the residual.
    The model captures the rise out of the underdoped regime and the plateau near optimal doping.}
    \label{fig:re123_dome}
\end{figure}

We collect all $RE$-123 entries in our dataset with $T_c > 0$ and normalize oxygen content to three Cu atoms per formula unit ($6+x \equiv 3\,\text{O}/\text{Cu}$) to place off-stoichiometric entries on a common axis. This yields 49 entries ($6.41 \le 6+x \le 7.10$, $19\,\text{K} \le T_c \le 95\,\text{K}$), of which 10 belong to our test set (Fig.~\ref{fig:re123_dome}). Without any explicit mechanism-specific inputs, $\mathcal{GP}$-$T_c$ successfully tracks the non-monotonic superconducting dome, reproducing the steep rise out of the underdoped regime and the 90--95~K plateau near optimal doping. The model achieves a mean absolute error of 8.3~K across all ten test entries, and 5.8~K across the seven entries in the optimal-doping regime ($6+x > 6.85$). This confirms that the graphlet histogram features subtly encode the local structural and bonding shifts that govern $T_c$ even in strongly correlated, doped unconventional superconductors.

\end{document}